\documentclass[a4paper,fleqn,usenatbib,useAMS]{mnras}

\usepackage{graphicx}	
\usepackage{float}
\usepackage{amsmath}	
\usepackage{amssymb}	
\usepackage{color}

\defcitealias{Ciarlariello2015}{CCP15}

\title[Modelling intrinsic size and luminosity correlations]{Modelling the impact of intrinsic size and luminosity correlations on magnification estimation}

\author[S. Ciarlariello and R. Crittenden]{
Sandro Ciarlariello,$^{1}$\thanks{E-mail: sandro.ciarlariello@port.ac.uk}
Robert Crittenden$^{1}$
\\
$^{1}$Institute of Cosmology and Gravitation, University of Portsmouth, Dennis Sciama Building, Portsmouth, PO1 3FX, UK\\
}

\date{Accepted XXX. Received YYY; in original form ZZZ}

\pubyear{2015}

\begin{document}
\label{firstpage}
\pagerange{\pageref{firstpage}--\pageref{lastpage}}
\maketitle

\begin{abstract}
Spatial correlations of the observed sizes and luminosities of galaxies can be used to estimate the magnification that arises through weak gravitational lensing. However, the intrinsic properties of galaxies can be similarly correlated through local physical effects, and these present a possible contamination to the weak lensing estimation. 
In an earlier paper \citep{Ciarlariello2015} we modelled the intrinsic size correlations using the halo model, assuming the galaxy sizes reflect the mass in the associated halo.  Here we extend this work to consider galaxy magnitudes and show that these may be even more affected by intrinsic correlations than galaxy sizes, making this a bigger systematic for measurements of the weak lensing signal.
We also quantify how these intrinsic correlations are affected by sample selection criteria based on sizes and magnitudes. 
\end{abstract}




\section{Introduction}
Weak gravitational lensing can be observed through the statistical analysis of coherent distortions in the shape, size and brightness of the images of distant galaxies. 
Measurements of galaxy shape correlations induced by weak lensing, also called cosmic shear, have been demonstrated to be a powerful probe and can potentially constrain the cosmological model with high precision.
Cosmic shear correlations were detected for the first time in 2000 \citep{Bacon2000, Kaiser2000a, vanWaerbeke2000, Wittman2000} and recently more accurately measured by surveys such CFHTLens \citep{Heymans2013} and KiDS \citep{Kuijken2015a}. Future surveys such as LSST\footnote{\tt www.lsst.org} and Euclid\footnote{\tt sci.esa.int/euclid} are expected to significantly improve shear measurements.
 
Although cosmic shear has traditionally been the primary goal of weak lensing studies, more attention has recently been given to size and brightness magnification as complementary probes. 
Magnification can push small or faint objects above the size and magnitude thresholds of a survey; this leads to a signal that can be detected by cross-correlating a foreground population of galaxies with a distant background sample. This is also known as magnification bias and was first detected by \cite{Scranton2005} using background quasars. More recently, other background sources such as Lyman-break galaxies have been used to study dark matter halo profiles \citep{Hildebrandt2009, vanWaerbeke2010a, Hildebrandt2011, Hildebrandt2013, Ford2012, Bauer2014}.

In \cite{Vallinotto2011} size and magnitude magnification were used to calibrate the cosmic shear measurements errors, but the first detection of cosmic magnification directly using galaxy sizes and magnitudes was performed in \cite{Schmidt2012}, where a weighted magnification estimator was applied to an X-ray-selected sample of galaxy groups; they found measurements of the projected surface density that are consistent with shear measurements.
\cite{Huff2014} used the Fundamental Plane relation for early-type galaxies to detect cosmic magnification by means of size measurements; however recently \cite{Joachimi2015a} detected a possible contamination from spatial correlations of Fundamental Plane residuals that should be taken into account. Recently, \cite{Duncan2016} presented the first measurement of individual cluster estimates using weak lensing size and flux magnification.

There are several good reasons for using size and magnitude information along with cosmic shear. Size and magnitude information is already available from cosmic shear surveys, and ideally one should exploit all of the data's statistical power to constrain the cosmological model.
For example, \cite{Casaponsa2013} have shown that the size information that comes from shape estimation methods can readily be used for cosmic magnification measurements, provided that there is sufficient signal-to-noise and the sizes are larger than the point spread function.
Using different weak lensing probes can be important to mitigate the impact of shape distortion systematics. 
For cosmic shear measurements, in addition to systematics arising from instrumental effects and atmospheric conditions, there are also systematics which have an astrophysical origin such as intrinsic alignments. The mechanisms that generate the intrinsic alignments are not fully understood and seem to depend on the galaxy type. The large-scale gravitational field seems to have a central role in generating alignments; essentially, the gravitational tidal field changes the shape of the halo in which an elliptical galaxy is embedded or, for spiral galaxies, it can induce angular momentum correlation to align their disc spins. These intrinsic alignments produce a signal that can mimic the effect of weak lensing and bias cosmological parameter constraints. A review of the various intrinsic alignment models and the methods to assess the contamination can be found in \cite{Troxel2015}. 

\cite{Heavens2013} have shown that adding cosmic magnification via size distortions can help to increase the constraining power compared to cosmic shear measurements on their own; they also demonstrated that size measurements can be made largely uncorrelated with shape measurements if the square root of the area of the galaxy image is used as size estimator.  
This analysis has been extended in \cite{Alsing2015} which provided an estimate of the convergence dispersion expected from size measurements and, analogously to intrinsic alignments for cosmic shear, they study the possible impact of marginalising over intrinsic size correlations on constraints of  cosmological parameters such as the dark energy equation of state parameters.
In \cite{Ciarlariello2015} (hereafter CCP15), we investigated a theoretical model in which intrinsic size correlations arise in a simple halo model, assuming larger and more massive galaxies reside in more massive haloes and linking observed galaxy sizes to halo and subhalo masses through the relation found by \cite{Kravtsov2013}. Haloes are populated with subhaloes by means of a subhalo mass function which accounts for the fact the size of the largest subhaloes is limited by the total halo mass. The main result from \citetalias{Ciarlariello2015} is that it may not be possible to ignore intrinsic correlations when weak lensing is measured from galaxy sizes.
 
In this paper we extend the analysis given in \citetalias{Ciarlariello2015} to account for intrinsic correlations of magnitudes. The halo model developed for intrinsic size correlations is applied to magnitudes and galaxy luminosities are correlated with the mass of the haloes and subhaloes following the relation given in \cite{Vale2008}.
In order to calculate the potential impact of these correlations on more realistic surveys we also include in our model size and magnitude thresholds.

The paper is organised as follows. In Section 2 we discuss how the convergence is estimated from size and magnitude measurements, in both the ideal case and for more realistic surveys.  Section 3 explains our halo model-based approach, how we relate the sub-halo masses to observed quantities and our model of the size-magnitude distribution.  Section 4 works out the relevant three dimensional two-point power spectra for the convergence and the intrinsic size and magnitude fields.  In Section 5 we translate these statistics for the two dimensional size and magnitude estimators, both for a fully projected sample and for a tomographic binning approach; we end with brief conclusions in Section 6.

\section{Magnification estimators}
\label{lensing-sec}

\subsection{Ideal size and magnitude estimators}
Weak gravitational lensing by large scale structure can be described by the Jacobian matrix that maps the true galaxy source positions to their observed positions on the sky,
\begin{equation}
A(\bmath\theta) = \left(\delta_{ij} - \frac{\partial^{2}\psi(\bmath\theta)}{\partial\theta_{i}\partial\theta_{j}} \right) =  \left(\begin{array}{cc} 1-\kappa-\gamma_{1} & -\gamma_{2} \\ -\gamma_{2} & 1-\kappa+\gamma_{1} \end{array} \right)\;,
\label{lensmatrix}
\end{equation}
where $\psi(\bmath\theta)$ is the two-dimensional gravitational potential, $\gamma_{1}= \frac{1}{2}(\psi_{,11}-\psi_{,22})$, $\gamma_{2} = \psi_{,12}$ where the comma in $\psi_{,i}$ is the partial derivative of the gravitational potential with respect to the variable $\theta_{i}$ and $\kappa$ is called the convergence. Indicating the cosmic shear by $\gamma$, we have: $\gamma = \gamma_{1} +\rm{i}\gamma_{2}$.
The determinant of this matrix gives the cosmic magnification $\mu$ of a surface area element:
\begin{equation}
\mu = \frac{1}{\det{A}}=[(1-\kappa)^2-|\gamma|^2]^{-1}\;.
\label{magnification}
\end{equation}
In the weak lensing regime $|\kappa|$ and $|\gamma|\ll 1$, so the magnification is approximately $\mu \simeq 1 + 2\kappa$.
Therefore, in the weak lensing limit, the observed galaxy sizes and fluxes, $r$ and $F$, are related to their intrinsic values by: 
\begin{equation}\begin{split}
r_{\rm{O}} &=(1+\kappa)r_{\rm{I}}\, \\
F_{\rm{O}} &= (1+2\kappa)F_{\rm{I}}\,
\end{split}
\label{sizefluxmagnification} 
\end{equation}
where the subscripts stand for the observed $(\rm O)$ and intrinsic $(\rm I)$ quantities. 
As pointed out by \cite{Heavens2013}, if galaxy size defined as the square root of the galaxy image it is expected to be uncorrelated with shear for galaxies with exponential profiles.
In order to get an estimator for the lensing convergence, we define the logarithm of the galaxy size in arcseconds and use the definition of apparent magnitude for galaxy fluxes, following \cite{Schmidt2012}, as follows:
\begin{equation}\begin{split}
\lambda &= \ln{\frac{r}{\rm [arcsec]}}\, \\
m &= m_{\rm ref} - 2.5\log_{10}\frac{F}{F_{\rm ref}}\,
\end{split}
\end{equation}
where $\lambda$ is logarithm galaxy size, $m$ is the galaxy magnitude and $m_{\rm ref}$ is the magnitude for a reference flux.
We then can use as our point estimators the following \citep{Schmidt2012, Heavens2013, Bacon2014}:
\begin{equation}\begin{split}
\hat{\kappa}^{\rm size} &= \lambda_{\rm{O}} - \langle \lambda_{\rm{O}}\rangle\,\\
\hat{\kappa}^{\rm mag} &= \frac{1}{q} (m_{\rm O}-\langle m_{\rm O} \rangle)\,,
\end{split}
\end{equation}
where  $q\simeq-2.17$.  

For any given galaxy, its observed size and magnitude will be determined more by its intrinsic values than by its magnification, so any individual measurement will be dominated by this intrinsic dispersion.  But by averaging many such measurements over a patch where the magnification is coherent, one can reach a regime where the magnification dominates.  
However, this assumes that the average intrinsic sizes and magnitudes are uncorrelated; if there are intrinsic correlations, so that 
$\langle r \rangle_{\rm patch} \ne \bar{r} $ and $\langle m \rangle_{\rm patch} \ne \bar{m} $  then this could be wrongly interpreted as magnification. 

The magnification estimator using either sizes or magnitudes will effectively have two contributions, the true lensing convergence and the intrinsic contribution: 
\begin{equation}\begin{split}
\hat{\kappa}^{\rm size} &= \kappa + \kappa_{\rm{I}}^{\lambda}\;, \\
\hat{\kappa}^{\rm mag} &= \kappa + \kappa_{\rm{I}}^{\rm m}\;.
\end{split}
\end{equation}
Here, $\kappa_{\rm{I}}$ is the intrinsic contribution to the magnification estimator arising from the intrinsic galaxy sizes and magnitudes.

\subsection{Incorporating selection effects}

So far we have focused on estimating lensing from an ideal survey, implicitly assuming that the population of objects is not affected by lensing.  In realistic surveys, the galaxies are only included if they exceed some thresholds for detection, either in magnitude, size or both.  In such a case, magnification can bring new objects into the survey, affecting the number density of objects and their average sizes and magnitudes.  This effect is generally called magnification bias \citep{Schmidt2009a}. 

Incorporating these effects, the average properties of the galaxies that enter into a sample are assumed to depend on the convergence as  
\begin{equation}\begin{split}
\langle\lambda_{\rm O}\rangle &= \langle\lambda_{\rm I}\rangle + \eta_{\lambda}\kappa \\
\langle m_{\rm O}\rangle &= \langle m_{\rm I}\rangle+ \eta_{m}\kappa.
\end{split}\label{respestimator}
\end{equation}
The quantities $\eta_{\lambda}$ and $\eta_{m}$ are called lensing responsivities \citep{Alsing2015}; in an ideal case (a survey with no cuts) we would have $\eta_{\lambda}=1$ and $\eta_{m}=q$. In general, these values will depend on both the survey population and observational thresholds and can be redshift dependent, e.g. \cite{Alsing2015} and \cite{Schmidt2012}. 
Below we show how the mean values and, consequently, the responsivities $\eta_{\lambda}$ and $\eta_{m}$ are changed when dealing with the realistic case of a survey with size and magnitude limits.

\subsubsection{Magnification bias: galaxy number density}

Here we briefly discuss the effect of magnification bias on the galaxy number density, following the treatment of \citet{Hui2007} and \citet{Schmidt2009a}; we then extend this to the properties relevant for convergence estimation using sizes and magnitudes.  
When there is either a magnitude or size threshold, magnification increases the number density as objects are brought in the sample;  at the same time, the solid angle is stretched leading to a dilution of the number density.
Which of these competing effects wins out depends on the density of galaxies at the edge of the selection cuts, but generically they cause correlations to be induced between the distant background galaxies (or quasars) and the foreground galaxies that trace the lensing potential.
 
We denote the observed and intrinsic distributions of galaxies as functions of log-sizes, magnitudes and positions by  $\Phi_{\rm O}(\lambda_{\rm O}, m_{\rm O}, \bmath\theta_{\rm O})$ and $\Phi_{\rm I}(\lambda_{\rm I}, m_{\rm I}, \bmath\theta_{\rm I})$ respectively.  
Conservation of the total number of galaxies implies:
\begin{equation}
d^2\theta_{\rm I}\,d\lambda_{\rm I}\,dm_{\rm I}\,\Phi(\lambda_{\rm I}, m_{\rm I}, \bmath\theta_{\rm I}) = d^2\theta_{\rm O} d\lambda_{\rm O} dm_{\rm O} \Phi_{\rm O}(\lambda_{\rm O}, m_{\rm O}, \bmath\theta_{\rm O}),
\end{equation}
where the relations below describe the change of size, magnitudes and area after a lensing transformation in an ideal case:
\begin{equation}\begin{split}
\lambda_{\rm O} &= \lambda_{I} + \kappa \\
m_{\rm O} &= m_{\rm I}  + q\kappa \\
d^2\theta_{\rm O} &= (1+2\kappa)d^2\theta_{\rm I}.
\end{split}
\end{equation}
The number density of objects in a survey is given by, 
\begin{equation}
n_{\rm O}(\bmath\theta_{\rm O}) = \int d\lambda_{\rm O}dm_{\rm O}\Phi_{\rm O}(\lambda_{\rm O}, m_{\rm O}, \bmath\theta_{\rm O})S(\lambda_{\rm O}, m_{\rm O}),
\end{equation}
where $S(\lambda, m)$ denotes the selection function of the survey.  For simplicity we assume the selection function to be spatially constant and a step function describing magnitude and size limits ($m_{\rm lim}$ and $\lambda_{\rm lim}$):
\begin{equation}
S(\lambda, m) = \Theta(\lambda - \lambda_{\rm lim})[1-\Theta(m-m_{\rm lim})]\,
\end{equation}
where the function $\Theta(x)$ is the Heaviside function. 

Using these relations, we can Taylor-expand the observed selection function, $S(\lambda_{\rm O}, m_{\rm O})$, with respect the convergence to find:
\begin{equation}\begin{split}
n_{\rm O}(\bmath\theta_{\rm O}) = (1\,-\,2\kappa)\int d\lambda_{\rm I}\,dm_{\rm I}\,&\Phi_{\rm I}(\lambda_{\rm I}, m_{\rm I}, \bmath\theta_{\rm I}) \\
&\times \left[S(\lambda_{\rm I}, m_{\rm I}) + \frac{\partial S}{\partial \lambda_{\rm I}}\kappa + q \frac{\partial S}{\partial m_{\rm I}}\kappa\right].
\end{split}
\end{equation}
If the function $S(\lambda_{\rm I}, m_{\rm I})$ is taken to be a step function, its derivatives are Dirac delta functions of either size or magnitude. Finally, we obtain:
\begin{equation}
n_{\rm O}(\bmath\theta_{\rm O}) = n_{\rm I}(\bmath\theta_{\rm I})[1 + (\zeta_{1} +\zeta_{2} - 2)\kappa(\bmath\theta_{\rm I})],
\end{equation}
where:
\begin{equation}\begin{split}
\zeta_{1} &= \int_{-\infty}^{m_{\rm lim}} dm_{\rm I}\,f_{\rm I}(\lambda_{\rm lim},m_{\rm I}) \\ 
\zeta_{2} &= -q\,\int_{\lambda_{\rm lim}}^{+\infty} d\lambda_{\rm I}\,f_{\rm I}(\lambda_{\rm I}, m_{\rm lim}).
\end{split}
\end{equation}
and we have defined:
\begin{equation}
f_{\rm I}(\lambda_{\rm I}, m_{\rm I}) \equiv \frac{\Phi_{\rm I}(\lambda_{\rm I},m_{\rm I})}{n_{\rm I}}.
\label{phinorm}
\end{equation}
These responsivities are defined in terms of the intrinsic galaxy properties, which we model directly here; however in practice the intrinsic properties will need to be inferred from the observed galaxy properties, which may introduce further uncertainties.

\subsubsection{Magnification bias: mean size and magnitude}

By means of the galaxy number density results from the previous section, we can calculate how mean values for sizes and magnitudes are affected by magnification bias. Eventually, we will obtain the responsivities when selection cuts are used in a survey.

In the following we begin with the calculation for the mean size. Analogous results for magnitudes follow the same reasoning.
The observed mean log-size is a region of the sky is given by:
\begin{equation}\begin{split}
\langle\lambda_{\rm O}\rangle(\bmath\theta_{\rm O}) = \frac{1}{n_{\rm O}} \int d\lambda_{\rm O}dm_{\rm O} \, \lambda_{\rm O} \Phi(\lambda_{\rm O},\,m_{\rm O},\,\bmath\theta_{\rm O}) S(\lambda_{\rm O}, m_{\rm O}).
\end{split}
\end{equation}
Translating everything into intrinsic quantities and accounting for the magnification bias effect in the galaxy number density, we obtain:
\begin{equation}\begin{split}
\langle\lambda_{\rm O}\rangle(\bmath\theta_{\rm O}) &= \frac{(1 -2\kappa)}{n_{\rm I}[1 + (\zeta_{1} +\zeta_{2} - 2)\kappa]}\int d\lambda_{\rm I}\,dm_{\rm I}\,\Phi_{\rm I}(\lambda_{\rm I}, m_{\rm I}, \bmath\theta_{\rm I}) \\
&\times(\lambda_{\rm I}+\kappa)\,\left[S(\lambda_{\rm I}, m_{\rm I}) + \frac{\partial S}{\partial \lambda_{\rm I}}\kappa + q\frac{\partial S}{\partial m_{\rm I}}\kappa\right].
\end{split}
\end{equation}
Carrying on the calculations, neglecting second order terms, we find:
\begin{equation}\begin{split}
\langle\lambda_{\rm O}\rangle(\bmath\theta_{\rm O}) = & \langle\lambda_{\rm I}\rangle(\bmath\theta_{\rm I}) + \kappa(\bmath\theta_{\rm I}) +(\alpha_{1} + \alpha_{2})\kappa(\bmath\theta_{\rm I})+ \\ 
&-\langle\lambda_{\rm I}\rangle(\zeta_{1}+\zeta_{2})\kappa(\bmath\theta_{\rm I})\,,
\end{split}
\end{equation}
where:
\begin{equation}\begin{split}
\alpha_{1} &= \int_{-\infty}^{m_{\rm lim}} dm_{\rm I}\,f_{\rm I}(\lambda_{\rm lim},m_{\rm I})\lambda_{\rm lim} \\
\alpha_{2} &= -q\,\int_{\lambda_{\rm lim}}^{+\infty} d\lambda_{\rm I}\,f_{\rm I}(\lambda_{\rm I}, m_{\rm lim})\lambda_{\rm I} .
\end{split}
\end{equation}
Finally, recalling eq. (\ref{respestimator}), we obtain the equation for the size responsivity:
\begin{equation}
{\eta}_{\lambda} = 1 + (\alpha_{1} + \alpha_{2}) - \langle\lambda_{\rm I}\rangle(\zeta_{1} + \zeta_{2}).
\label{etatildelambda}
\end{equation}

The results for the magnitude responsivity are very similar:
\begin{equation}
{\eta}_{\rm m} = q + (\beta_{1} + \beta_{2}) - \langle m_{\rm I}\rangle(\zeta_{1} + \zeta_{2}),
\label{etatildemag}
\end{equation}
where we have:
\begin{equation}\begin{split}
\beta_{1} &= \int_{-\infty}^{m_{\rm lim}} dm_{\rm I}\,f_{\rm I}(\lambda_{\rm lim},m_{\rm I})m_{\rm I} \\
\beta_{2} &= -q\,\int_{\lambda_{\rm lim}}^{+\infty} d\lambda_{\rm I}\,f_{\rm I}(\lambda_{\rm I}, m_{\rm lim})m_{\rm lim}.
\end{split}
\end{equation}

The results described above are for a given population and convergence. It is worth remembering that the convergence field is redshift dependent, and in addition that the characteristics of the populations of galaxies change with redshift.  
In particular,  
\begin{equation}\begin{split}
\langle\lambda_{\rm O}(z)\rangle - \langle\lambda_{\rm I}(z)\rangle &= \eta_{\lambda}(z)\kappa (z)\\
\langle m_{\rm O}(z)\rangle - \langle m_{\rm I}(z)\rangle & =  \eta_{m}(z)\kappa(z) .
\end{split}
\end{equation}
For example, a low redshift population of galaxies is likely to see small levels of magnification; however, since most of the population will already be in the sample, it will also see nearly ideal responsivities.  In contrast, at high redshifts the expected convergence level is much higher, but the responsivities are likely to be lower as more galaxies are likely to be below the observational thresholds.  

\section{Modelling intrinsic correlations}

Here we briefly describe a model to account for intrinsic correlation of sizes and magnitudes, based on the halo model formalism \citep{Scherrer1991,Seljak2000,Cooray2002,Sheth2003}. 
The basic assumption of the halo model is that the mass in the Universe is distributed in distinct haloes. In our model, we assume that a central galaxy is associated with the halo centre, and satellite galaxies are distributed around it following a profile probability density. The satellites are associated with sub-haloes, which have a mass distribution that depends on the mass of the halo in which they sit. 
Furthermore, we assume the central galaxy is has a mass given by the total halo mass minus the mass in sub-haloes hosting satellite galaxies. 
Throughout we indicate halo masses and sizes with $M$,$R$ and sub-halo (or satellite) masses and sizes with $m_{\rm sh}$, $r$; the mass associated with the central galaxy is given by $M_{\rm c} = M - \Sigma_i m_{\rm{sh},i}$.

\subsection{Elements of the halo model} 
\label{halomodel-sec}

A complete specification of the halo model requires knowing the halo mass function and the distribution of sub-halo masses within a halo; it also requires knowing the probability density profile of how sub-haloes are distributed in a halo and understanding the statistics of how haloes are distributed on large scales, usually parameterised by the mass dependent bias function.  
We now briefly recall these main features; we use the same notation and equations provided in \citetalias{Ciarlariello2015} and refer the reader there for more details regarding our halo model assumptions.

In this paper, we use the halo mass function $n_{\rm com}(M,z)$ given by \cite{Sheth1999}, describing the comoving number density of collapsed haloes; the distribution of satellite galaxies is described by the sub-halo mass function $dN(m_{\rm sh}|M,z)/dm_{\rm sh}$ from \cite{Giocoli2010a}, which depends on the host halo mass and accounts for the fact that more massive sub-haloes only exist in more massive haloes.  We also assume that sub-haloes are distributed around the centre of the halo according to a Navarro-Frenk-White (NFW) profile \citep{Navarro1996}; in particular, since we work with power spectra, we use the Fourier transform of the NFW profile normalized to the halo mass indicated with $u(k|M)$. In principle, we should also specify a density profile for sub-haloes, as in \cite{Giocoli2010a}; however we assume simple relations of satellite radii and luminosities with the sub-halo mass, so the sub-halo profiles are not required. 

Finally, we need to describe the large-scale distribution of haloes; this is usually done by specifying the two-point moments and we assume a simple deterministic bias that is mass dependent.
In the halo model, the two-point correlation function can be written 
\begin{equation}
\xi(\bmath{x}) = \xi_{\rm{1h}}(\bmath{x}) + \xi_{\rm{2h}}(\bmath{x})\;,
\end{equation}
where the first term describes the contribution from each halo whereas the second term gives the contribution on large scales from halo correlations. The mass function and probability density profiles are needed to evaluate both terms, but the two-halo term also requires the halo correlation function $\xi_{\rm{hh}}(\bmath{x}|M_1, M_2)  = b(M_{1})b(M_{2})\xi_{\rm{lin}}{(\bmath{x})}$ where $\xi_{\rm{lin}}(\bmath{x})$ is the linear mass correlation function and $b(M, z)$ is the bias parameter.   
We use the bias model (consistent with the mass function) from \cite{Sheth1999}.
Moreover, as pointed out by \cite{Seljak2000}, because on large scale the amplitude of the two-halo term has to match the amplitude of the linear power spectrum, we have a constraint for the halo model bias so that, on the very largest scales where the mass profile of the haloes is unimportant, the mass distribution matches linear theory.  
In the following sections the redshift dependence of the halo mass function and the halo bias is always implicitly assumed even though it is not indicated.

\subsection{From haloes to galaxies}
\label{halogal}

We require a process for relating the sizes and magnitudes of galaxies to the halo and sub-halo masses in the halo model. 
For this, we use relations found by \cite{Kravtsov2013} and by \cite{Vale2008} to relate halo masses to galaxy sizes and luminosities repectively. 
Both of these relations were found by means of abundance matching, which relates simulated halo masses to the properties of observed galaxies.

By this means \cite{Kravtsov2013} found a linear relation between the virial radius $R_{200}$ of the haloes and radius enclosing half of the galaxy mass $r_{1/2}$ holding over eight orders of magnitude in stellar mass and for all morphological types:
\begin{equation}
r_{1/2} = 0.015 R_{200}.
\label{halfmassrad}
\end{equation}
In this work we use the effective radius $R_{\rm e}$ which is defined as the radius in which half of the light of the galaxy image is contained and is simply related to $r_{1/2}$ through the relation given in \cite{Kravtsov2013}:
\begin{equation}
R_{\rm e}=\frac{r_{1/2}}{1.34}. 
\end{equation}
In the following we identify $R_{\rm e}$ with $r(m_{\rm sh})$ in order to keep the notation concise.

Regarding luminosities, \cite{Vale2008} used abundance matching to fit the mass-luminosity relation for individual galaxies, using the following double power-law:
\begin{equation}
L = L_0\frac{(M/M_0)^a}{[1+(M/M_0)^{bk}]^{1/k}}.
\label{fitL}
\end{equation}
In principle abundance matching can be used for any choice of waveband, provided the luminosity function is well-constrained.
\cite{Vale2008} found for the K waveband, $L_0=1.37\times 10^{10}\,L_\odot\,h^{-2}$, $M_0=6.14\times 10^{9}\,M_\odot\,h^{-1}$, $a=21.03$, $b=20.74$ and $k=0.0363$, while for the B$_{\rm j}$ waveband the parameters are: $L_0=4.12\times 10^{9}\,L_\odot\,h^{-2}$, $M_0=1.66\times 10^{10}\,M_\odot\,h^{-1}$, $a=6.653$, $b=6.373$ and $k=0.111$ \citep{Vale2008}.
In the following, we primarily assume our galaxy luminosities are provided in the $Bj$ waveband, as these provide the more conservative results, but we also provide the size-magnitude probability distribution in the $K$ waveband for comparison. 

Apparent and absolute magnitudes ($M_{\rm abs}$) are related by: 
\begin{equation}
m - M_{\rm abs} = 25 + 5\log_{10}\frac{d_L}{[{\rm Mpc}]},
\end{equation}
and the absolute magnitudes are defined as,
\begin{equation}
M_{\rm abs} - M_{{\rm abs},\odot} = -2.5\log_{10}\frac{L}{[L_{\odot}]},
\end{equation}
where we need to remember that each quantity is defined in a certain waveband. In particular, $M_{{\rm abs},\odot}$ is the solar absolute magnitude in a well defined waveband.
Finally we obtain the equation which can link apparent magnitudes with luminosities and, eventually, halo masses:
\begin{equation}
m = 25 + 5\log_{10}{\frac{d_{L}(z)}{\rm [Mpc]}} - 2.5\log_{10}{\frac{L}{[L_{\odot}]}} + M_{\rm{abs},\odot}\,,
\label{minmag}
\end{equation}
where $M_{\rm{abs},\odot}$ is the solar absolute magnitude in the chosen waveband.

\subsection{Modelling the size-magnitude distribution}
\label{sizemagpdf}

\begin{figure*}
\centering
\includegraphics[scale=0.6]{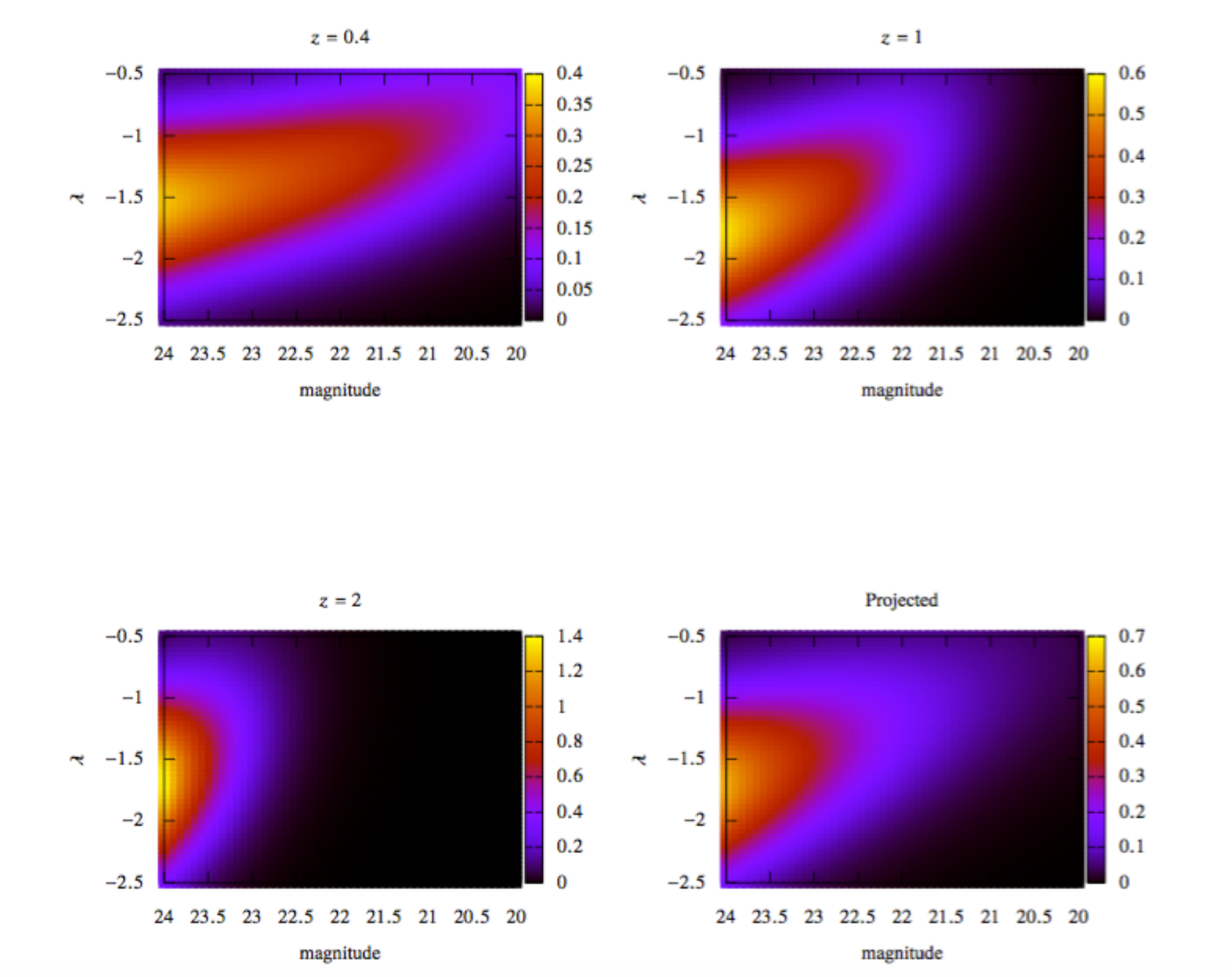}
\caption{The inferred size-magnitude distribution from our halo model in Bj-band for different redshifts and projected over a Euclid-like survey. Cuts are m$_{\rm lim}=24$ and $\lambda_{\rm lim}=-2.5$.}
\label{fig:MATRIX_Bj}
\end{figure*}

\begin{figure}
\centering
\includegraphics[scale=0.45]{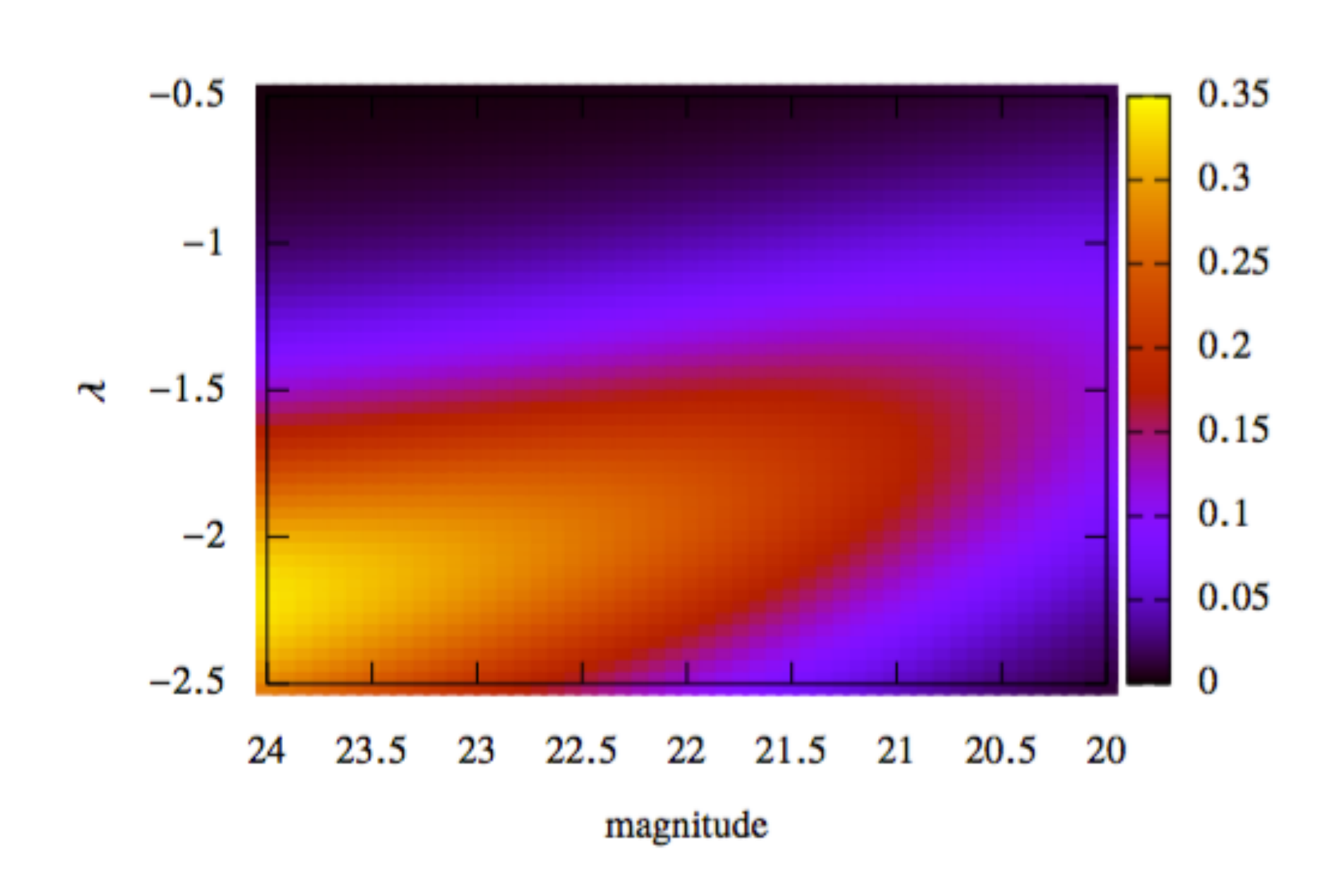}
\caption{The inferred size-magnitude distribution in K-band, projected over a Euclid-like survey. Cuts are m$_{\rm lim}=24$ and $\lambda_{\rm lim}=-2.5$.}
\label{fig:MATRIX-091_K}
\end{figure}

In our model, the intrinsic size and magnitude distribution is essentially given by the integral of the halo and sub-halo mass functions.  We use the physical halo mass function corrected for the co-moving volume:
\begin{equation}\begin{split}
n(M,z) &= \frac{dN}{dMdzd\Omega} = \frac{dN}{dMdV_{\rm com}}\frac{dV_{\rm com}}{dzd\Omega} \\
&= n_{\rm com}(M,z)\frac{c}{H_{0}}\frac{(1+z)^2\,D_{A}^2(z)}{E(z)}.
\end{split}
\end{equation}
Here, $D_A(z)$ is the angular diameter distance and $E(z)$ describes the evolution of the Hubble parameter. 

We also model the additional intrinsic scatter in the galaxy size-halo mass and luminosity-halo mass relations; in particular, we model $P(\lambda, m| m_{\rm sh}, z)$, the probability that a galaxy at a given mass and redshift are observed with a given size and magnitude. 
For simplicity we model this as a product of Gaussian distributions centred around the mean values described above, 
\begin{equation}
P(\lambda, m| m_{\rm sh}, z) = N(\lambda | \lambda(m_{\rm sh},z), \sigma_{\lambda})N(m | m(m_{\rm sh},z), \sigma_{m})
\label{eq:scatter} 
\end{equation} 
where 
\begin{equation}\begin{split}
N(\lambda | \lambda(m_{\rm sh},z), \sigma_{\lambda}) &= \frac{1}{\sqrt{2\pi}\sigma_{\lambda}}\exp\left(-\frac{(\lambda -\lambda(m_{\rm sh},z))^2}{2\sigma_{\lambda}^2}\right) \\ 
N(m | m(m_{\rm{sh}},z),\sigma_{m}) &= \frac{1}{\sqrt{2\pi}\sigma_{m}}\exp\left(-\frac{(m-m(m_{\rm sh},z))^2}{2\sigma_{m}^2}\right).
\end{split}
\end{equation} 
For the scatter, we use the values found by \cite{Kravtsov2013} for the size-virial radius relation of galaxies of $0.2 \,{\rm dex}$, corresponding to an intrinsic scatter of $\sigma_{\lambda} \simeq 0.46$. For the luminosity-halo mass relation we use the value found by \cite{Hansen2009}, that is $\sigma_{\log_{10}{L}} \sim 0.2$; translating into magnitudes, we find $\sigma_{m}=0.5$.
For the present, we assume that these scatters are uncorrelated and independent of redshift. 

The resulting size-magnitude distribution is given by: 
\begin{equation}\begin{split}
\Phi_{\rm I}(\lambda_{\rm I}, m_{\rm I},z) & = \int_{M_{*}}^{+\infty} dM\,n(M,z)  \\ &\times\int_{M_{*}}^{M}dm_{\rm sh}\,\left(\delta_{\rm D}(m_{\rm sh}-M_{\rm c}) + \frac{dN(m_{\rm sh}|M,z)}{dm_{\rm sh}}\right) \\
&\times N(\lambda_{\rm I}| \lambda(m_{\rm sh},z), \sigma_{\lambda})N(m_{\rm I}| m(m_{\rm sh},z), \sigma_{m}) \label{eq:Phidist}
\end{split}
\end{equation}
where $M_{*}=10^{10}\,M_\odot\,h^{-1}$ indicates the minimum mass for haloes hosting a galaxy (see \cite{Guo2010}) and $M_{\rm c}$ accounts for the fact that the central galaxy has somewhat less mass than the full halo.

In Fig. \ref{fig:MATRIX_Bj} the size-magnitude distributions in the Bj waveband are shown for different redshifts.  As expected, the observed sizes and fluxes are larger at lower redshifts.  Also plotted is the size-magnitude distribution integrated over redshift, which in shape resembles that of the mean redshift. 
In Fig. \ref{fig:MATRIX-091_K}, we also plot the integrated size-magnitude distribution in the K waveband, derived from the K-band luminosity-mass function given above. Results in those figure are shown for magnitudes between  $24<m<20$ and sizes between $-2.5<\lambda<-0.5$. We assume the same cuts throughout this work if not otherwise stated.

\subsection{Survey thresholds and responsivities}

To obtain more realistic results, we need to model the galaxy selection effects. In this work, we assume a magnitude-limited survey with $m_{\rm{lim}}$.   The limiting magnitude value refers to a given waveband and here we assume values for either for $B_j$ or $K$ waveband corresponding to the luminosity-halo mass relations given by \cite{Vale2008}.
It is also necessary to model the selection effect for galaxy sizes, and assume a limiting size of $r_{\rm min}$ in arc seconds (in practise for realistic observations, it might also be necessary to consider cuts on the galaxy surface brightness).

Given a model for the size-magnitude distribution and assuming a selection function, it is also possible to calculate the redshift distribution of the survey :
\begin{equation}
p(z) = \frac{\int_{\rm cuts} d\lambda\,dm\,\Phi(\lambda,m,z)}{\int dz\,\int_{\rm cuts} d\lambda\,dm\,\Phi(\lambda,m,z)},
\label{redshiftdistr}
\end{equation}
Assuming, for the Bj-band, sharp cuts at $\lambda_{\rm lim} = -2.5$ and $m_{\rm lim}=24$ our model can roughly reproduce the expected redshift distribution with mean redshift $\langle z \rangle \sim 0.96$.

\begin{figure}
\centering
\includegraphics[scale=0.45]{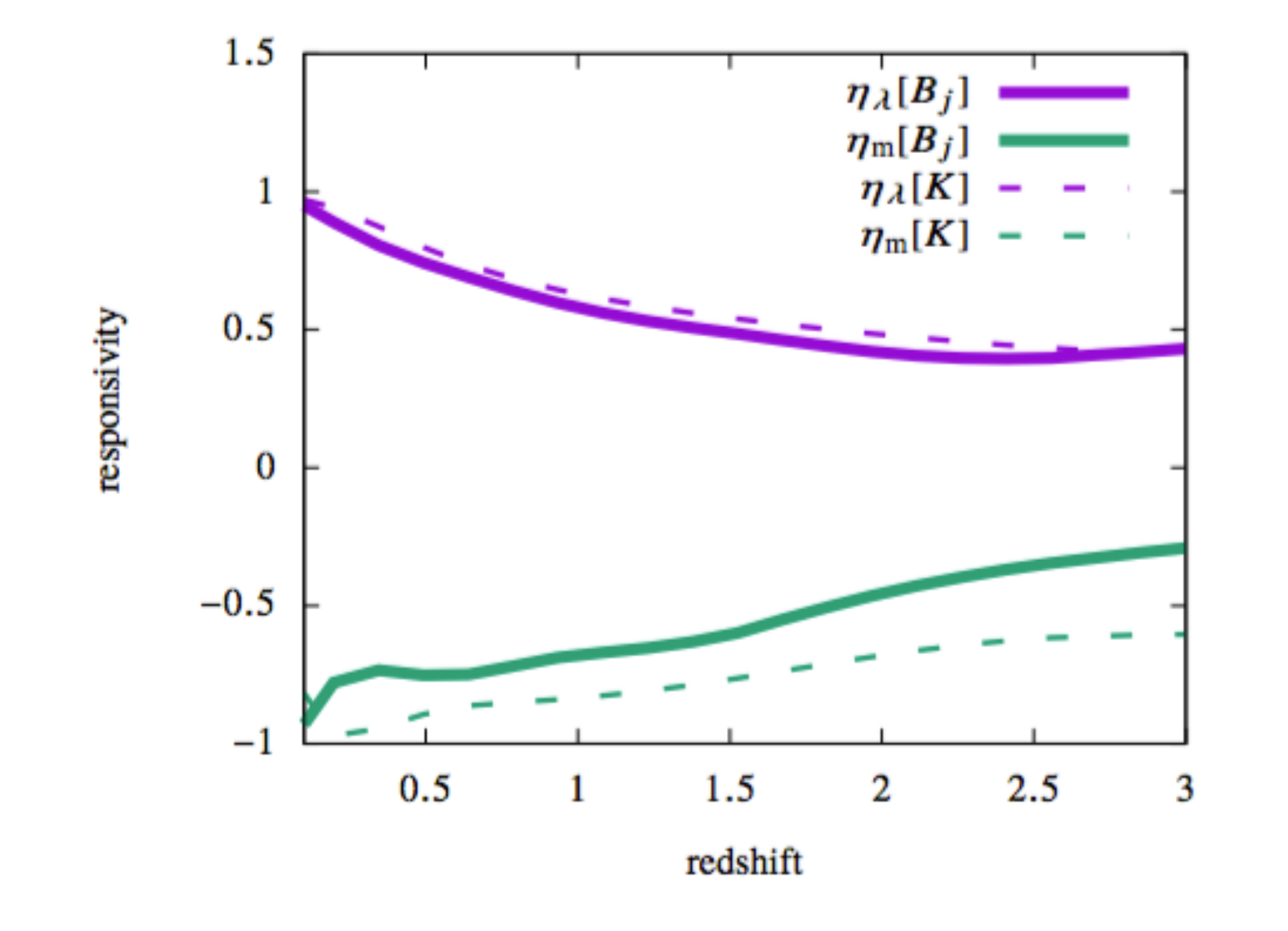}
\caption{Responsivity values for size and magnitude as a function of redshift.  Results for the $B_j$ band are solid and those for the $K$ band are dashed. }
\label{fig:responsivity}
\end{figure}

With these survey thresholds, we can derive the size and magnitude responsivities of the model as a function of redshift as described above.  
These are shown in  Fig. \ref{fig:responsivity}.   We can see that the the size responsivity approaches the ideal ($\eta_\lambda = 1$) at low redshifts; this reflects the fact that the magnitude cut, rather than the size cut, is of primary importance at these redshifts, as can be seen in Fig. \ref{fig:MATRIX_Bj}. 
The magnitude responsivity is significantly different from the ideal  ($\eta_m = -2.17$), reflecting the fact that many galaxies can be pulled into the sample by magnification.  

Note that in calculating the redshift evolution of the responsivity, we do not include `k-corrections'; our results in  Fig. \ref{fig:responsivity} simply account for the shrinking and dimming of galaxies seen at further distances, and not the fact that they are observed in a different rest-frame frequency, which would require assuming an average spectral shape for the galaxies.  However, in  Fig. \ref{fig:responsivity} we do show the responsitivies derived from mass-luminosity relations calibrated in different bands (Bj and K) \citep{Vale2008}.   

To calculate the average responsivity for a survey distributed over a wide redshift range, we must first integrate the size-magnitude distribution function over redshift.  As described above, the responsivities can be derived from:    
\begin{equation}
f_{\rm I}(\lambda_{\rm I}, m_{\rm I}) = \frac{\int dz\,\Phi(\lambda,m,z)}{\int dz\,\int_{\rm cuts} d\lambda\,dm\,\Phi(\lambda,m,z)}.
\end{equation}
Recall however, this is a mean responsivity and different redshifts respond differently; the redshift dependence re-weights the effective convergence as described below.   

\subsection{Conditional moments}  

As the statistical properties of galaxies in haloes depend on their mass, it is useful to understand how the mean properties of galaxies that are selected 
depend on their mass.   In particular, we are interested in the probability that a galaxy of a given mass and redshift enters into the sample, and how the mean sizes and magnitudes are affected.  
For selection thresholds $\lambda_{\rm{lim}}$ and $m_{\rm{lim}}$, the probability of observing a galaxy with sub-halo mass $m_{\rm sh}$ and at redshift $z$  is
\begin{equation}
P_{\rm obs} (m_{\rm sh}, z) =   \int_{\lambda_{\rm lim}}^{+\infty} d\lambda\,  \int_{-\infty}^{m_{\rm lim}} dm\,  P(\lambda, m| m_{\rm sh}, z),  
\end{equation}  
where again we use the Gaussian model for the scatter given in Eq. \ref{eq:scatter}. 
Similarly, the conditional moments, 
 $\tilde\lambda(m_{\rm sh},z)$ and $\tilde{m}(m_{\rm sh},z)$, can be defined as follows:
\begin{equation}\begin{split}
\tilde{\lambda}(m_{\rm sh}, z) &= \int_{\lambda_{\rm lim}}^{+\infty} d\lambda\,\int_{-\infty}^{m_{\rm lim}} dm\ \lambda\,   P(\lambda, m| m_{\rm sh}, z)  , \\
\tilde{m}(m_{\rm sh}, z) &=  \int_{\lambda_{\rm lim}}^{+\infty} d\lambda\,  \int_{-\infty}^{m_{\rm lim}} dm \, m \, P(\lambda, m| m_{\rm sh}, z) . 
\end{split}
\label{eq:meancuts}
\end{equation}
Note that these are unnormalised weighted moments.  If one was interested for example in the average size of galaxies with a given underlying mass that makes it through the observational cuts, this would be given by $\bar{\lambda} (m_{\rm sh}, z) = \tilde{\lambda}(m_{\rm sh}, z) / P_{\rm obs} (m_{\rm sh}, z)$.  

\section{Correlation statistics}

In this section we work out the one and two-point statistics of the size and magnitude fluctuation fields.  Note that below we focus on the intrinsic quantities, and so drop subscripts on the size and magnitude variables. 

\subsection{Galaxy size and luminosity fields}

For a particular realisation of a galaxy field in the halo model, we can define the discrete galaxy density field as an integral over the observed size-magnitude distribution:
 \begin{equation}\begin{split}
n_{\rm g}(\bmath{x}) =  & \int_{\lambda_{\rm lim}}^{+\infty} d\lambda\, \int_{-\infty}^{m_{\rm lim}}dm\,\sum_{i} \sum_{j} \\ &\times \delta_{\rm D}(\lambda-\lambda_j) \delta_{\rm D}(m-m_j) \delta_{\rm D}^{(3)}(\bmath{x}-\bmath{x}_{i}-\bmath{x}_{j}) 
 \end{split}
\end{equation}
where the $\sum_{i}$  is over the halos and the $\sum_{j}$ is over the central and possible satellite galaxies within each halo.  $ \bmath{x}_{j}$ represents the position of each galaxy relative to the halo centre, $\bmath{x}_{i}$; $ \bmath{x}_{j}=0$ for the central galaxy, while for the satellite galaxies, their positions are described by the satellite probability profile.  The limits of the size and magnitude integrations guarantee that we only include galaxies that pass the observational selection criteria. 

We can take the average of the discrete density to determine the mean densities at each redshift:
\begin{equation}
 \bar{n}_{\rm g} (z) = \int_{\lambda_{\rm lim}}^{+\infty} d\lambda\,\int_{-\infty}^{m_{\rm lim}}dm \, \Phi (\lambda, m, z). 
\end{equation} 
 Inserting the definition of $ \Phi (\lambda, m, z)$ from the halo model (eq. \ref{eq:Phidist}), we can see that after rearranging the order of the integrals, this can be written as 
\begin{equation}\begin{split}
 \bar{n}_{\rm g} (z) &= \int_{M_{*}}^{+\infty} dM\,n(M,z) \int_{M_{*}}^{M}dm_{\rm sh}\, \\ & \times \left(\delta_{\rm D}(m_{\rm sh}-M_{\rm c}) + \frac{dN(m_{\rm sh}|M,z)}{dm_{\rm sh}}\right)  P_{\rm obs}(m_{\rm sh}, z).  \end{split}
 \end{equation}
Here, we will largely follow \citetalias{Ciarlariello2015} in calculating the statistics of the size and magnitude estimators. 
As can be seen from the equation above, the main difference with our previous treatment is that instead of using a fixed mass threshold at all redshifts, 
the mass threshold is redshift dependent and is a tapering function of the sub-halo mass.   The latter accounts for the scatter in the mass-size and mass-magnitude relations, while the redshift dependence accounts for the fact that hard cuts in the observed sizes and magnitudes correspond to different physical sizes and luminosities at different redshifts.  

Similar arguments can be applied to the size and magnitude fields. 
We weight the galaxy density field by their sizes and magnitudes to define, respectively, the galaxy size and luminosity fields as:
 \begin{equation}\begin{split}
\lambda(\bmath{x}) &=  \bar{n}_{\rm g}^{-1} \int_{\lambda_{\rm lim}}^{+\infty} d\lambda\,\int_{-\infty}^{m_{\rm lim}}dm\, \sum_{i}\sum_{j}  \\ &\times \delta_{\rm D}(\lambda-\lambda_j) \delta_{\rm D}(m-m_j) \delta_{\rm D}^{(3)}(\bmath{x}-\bmath{x}_{i}-\bmath{x}_{j})\lambda_j \\
m(\bmath{x}) &=  \bar{n}_{\rm g}^{-1} \int_{\lambda_{\rm lim}}^{+\infty} d\lambda\,\int_{-\infty}^{m_{\rm lim}}dm\, \sum_{i} \sum_{j} \\ &\times \delta_{\rm D}(\lambda-\lambda_j) \delta_{\rm D}(m-m_j) \delta_{\rm D}^{(3)}(\bmath{x}-\bmath{x}_{i}-\bmath{x}_{j})m_j . 
\end{split}
\label{sizemagfield}
\end{equation}
Following arguments similar to those above, the mean sizes and magnitudes can be expressed in terms of $ \Phi (\lambda, m, z)$, and the integration orders swapped, leading to: 
\begin{equation}\begin{split}
\langle\lambda\rangle (z) &= \frac{1}{\bar{n}_{\rm g}}\int_{M_{*}}^{+\infty} dM\,n(M,z)\int_{M_{*}}^{M}dm_{\rm sh} \\ &\times\left(\delta_{\rm D}(m_{\rm sh}-M_{\rm c}) + \frac{dN(m_{\rm sh}|M,z)}{dm_{\rm sh}}\right)\tilde\lambda(m_{\rm sh},z) \\
\langle m \rangle (z)  &= \frac{1}{\bar{n}_{\rm g}}\int_{M_{*}}^{+\infty} dM\,n(M,z)\int_{M_{*}}^{M}dm_{\rm sh} \\ &\times\left(\delta_{\rm D}(m_{\rm sh}-M_{\rm c}) + \frac{dN(m_{\rm sh}|M,z)}{dm_{\rm sh}}\right)\,\tilde{m}(m_{\rm sh},z)\,.
\end{split}
\end{equation}
These expressions are very similar to the form used in \citetalias{Ciarlariello2015}, but replacing $\lambda(m_{\rm sh}) \rightarrow \tilde\lambda(m_{\rm sh}, z)$ and $m(m_{\rm sh}) \rightarrow \tilde{m}(m_{\rm sh})$.  Again, this change folds in the redshift dependent cut and the scatter in the mass-size and mass-magnitude relations. 
With this modification, the subsequent calculations of the statistics of the fields go forward as in \citetalias{Ciarlariello2015}.

\subsection{The local estimator fields}

We are interested in the statistics of the intrinsic fluctuations of the observed angular sizes and magnitudes away from their mean values at a given redshift or integrated over a range of redshifts, as these are what are used as estimators for the convergence. 
In particular, we define
\begin{equation}
\begin{split}
\delta_{\rm I}^{\lambda} (\hat{\theta}, z) = \, & \lambda (\hat{\theta},z) - \langle\lambda \rangle_{z} \\
\delta_{\rm I}^{m} (\hat{\theta}, z) = \, & m(\hat{\theta},z) - \langle m \rangle_{z} .
\end{split} 
\end{equation}
At a fixed redshift, we can link these directly to the physical sizes and luminosities of the galaxies.  

Let us focus on the size fluctuation field.   A galaxy of physical size $r$ at redshift $z$ has 
 \begin{equation}
\lambda  (\hat{\theta}, z)= \ln{\frac{r(\hat{\theta}, z)}{\rm [Mpc]}} - \ln{\frac{D_{\rm A}(z)}{\rm [Mpc]}}.
\end{equation}
All galaxies at this redshift experience the same offset, so that the angular diameter distance terms cancel, leaving  
\begin{equation}
\delta_{\rm I}^{\lambda}  (\hat{\theta}, z) = \ln{\frac{r (\hat{\theta}, z)}{\rm [Mpc]}} - \left\langle \ln{\frac{r}{\rm [Mpc]}} \right\rangle_{z}\;.  
\label{eqn:kI}
\end{equation}
The scatter in the observed angular sizes of a galaxy of fixed mass $m_{\rm sh}$ (at a fixed redshift) entirely arises from the scatter of the physical sizes
that can be associated with this mass. 
While irrelevant in terms of the fluctuations, the angular diameter distance is still relevant for determining whether a particular galaxy is in the survey.
As above, we denote the weighted contribution to the size fluctuation field that survives the selection cuts as $\tilde{\delta}_{\rm I}^{\lambda} (m_{\rm sh}, z)$.

We obtain equivalent relations for the magnitude fluctuation field and its relation to the galaxy luminosity at a given redshift.  In particular, 
we have:
\begin{equation}
\delta_{\rm I}^{\rm m}   (\hat{\theta}, z)= -2.5\log_{10}{\frac{L (\hat{\theta}, z)}{[L_{\odot}]}} +2.5 \left\langle \log_{10}{\frac{L}{[L_{\odot}]}} \right\rangle_{z}\;.  
\end{equation}
Again, we denote the weighted contribution to the magnitude fluctuation field that survives the selection cuts as $\tilde{\delta}_{\rm I}^{m} (m_{\rm sh}, z)$.

Regardless of whether we are dealing with the size or magnitude field, a given realization of halo and sub-halo positions results in an estimator-weighted density field as:
 \begin{equation}\begin{split}
\delta_{\rm I}(\bmath{x}) &= \bar{n}_{\rm g}^{-1} \int_{\lambda_{\rm lim}}^{+\infty} d\lambda\,\int_{-\infty}^{m_{\rm lim}}dm\, \sum_{i} \\ &\times \sum_{j} \delta_{\rm D}(\lambda-\lambda_j) \delta_{\rm D}(m-m_j) \delta_{\rm D}^{(3)}(\bmath{x}-\bmath{x}_{i}-\bmath{x}_{j})\delta_{\rm I,j}.
\end{split}
\end{equation}
This can be rewritten as, 
 \begin{equation}\begin{split}
\delta_{\rm I}(\bmath{x}) =& \, \bar{n}_{\rm g}^{-1} \int_{M_*}^{+\infty} dM \, \sum_{i}  \delta_{\rm D}(M-M_i) \int_{M_*}^{M} dm_{\rm sh}  \\ &\times \sum_{j}\delta_{\rm D}(m_{\rm sh}-m_j) \delta_{\rm D}^{(3)}(\bmath{x}-\bmath{x}_{i}-\bmath{x}_{j}) \tilde{\delta}_{\rm I} (m_{\rm sh}, z).
\end{split}
\end{equation}

\subsection{Auto-correlation power spectra}
Our focus here is to understand the implications of size correlations on two-point statistics, and in particular in comparing how the power spectrum of the magnification estimator relates to that of the true magnification once size correlations are included.  
Thus, we must calculate the power spectrum of the intrinsic size correlations and their cross correlation with the true magnification.  

 As discussed above, in the halo model two-point correlations receive contributions from pairs of galaxies inhabiting the same halo and from where they inhabit two different haloes.  
The same holds for the power spectrum:
\begin{equation}
P(k) = P_{\rm{1h}}(k) + P_{\rm{2h}}(k)\;.
\end{equation}
It is straightforward to calculate the power spectrum of the matter density fluctuation $\delta \rho/\bar{\rho}$  using the halo model formalism developed above , e.g \citep{Scherrer1991}, \cite{Seljak2000}, \cite{Sheth2003}.

In the following subsections we present our model for the correlation between intrinsic correlation of galaxy sizes and magnitudes. In the one-halo terms, we only include the cross-correlations between different galaxies, so there is no central-central contribution.  

As the expressions are very similar, we use $\delta_{\rm I}$ to indicate the intrinsic contributions from either sizes or magnitudes, $\delta_{\rm I}^{\lambda}$ or $\delta_{\rm I}^{\rm m}$.
In order to reflect the effects of the cuts in the size-magnitude distribution, we indicate the generic intrinsic field as $\tilde\delta_{\rm I}$ as discussed above.  Initially, we derive the three-dimensional power spectra at a fixed redshift, and so drop the explicit redshift dependences; we discuss in the next section projecting this into two dimensional correlation functions and tomographic analyses.  

\subsubsection{One-halo terms}

Applying the halo model formalism, we obtain the following power spectra for the satellite auto-correlation:
\begin{equation}\begin{split}
P^{\rm{1h-sat}}_{\delta_{\rm I}}(k) &= \bar{n}_{\rm g}^{-2} \int_{M_*}^{\infty} dM n(M)\\
&\times\left[\int_{M_{*}}^{M} dm_{\rm sh} \frac{dN(m_{\rm sh}| M)}{dm_{\rm sh}}\tilde\delta_{\rm I}(m_{\rm sh}) u_{\rm d}(k, M)\right]^{2} . \end{split}
\end{equation}
We also have contribution from central-satellite correlation term:
\begin{equation}\begin{split}
P^{\rm{1h-cen-sat}}_{\delta_{\rm I}}(k) &= \frac{2}{\bar{n}_{\rm g}^2} \int_{M_{*}}^{\infty} dM n(M)\,\tilde\delta_{\rm I}(M_{\rm c}) \\  &\times\int_{M_{*}}^{M} dm_{\rm sh} \frac{dN(m_{\rm sh}| M)}{dm_{\rm sh}} \tilde\delta_{\rm I}(m_{\rm sh}) u_{\rm d}(k, M) .
\end{split}
\end{equation}

\subsubsection{Two-halo terms}

Applying the halo model formalism, we obtain the following power spectra for the auto-correlation:
\begin{equation}\begin{split}
P^{\rm{2h}}_{\delta_{\rm I}}(k) = (\bar{b}_{\delta_{\rm{I, cen}}}+ \bar{b}_{\delta_{\rm{I, sat}}})^{2}P^{\rm{lin}}(k)\;, 
\end{split}
\end{equation}
where:
\begin{equation}
\bar{b}_{\delta_{\rm{I, cen}}} = \bar{n}_{\rm g}^{-1} \int_{M_{*}}^{\infty} dM\,n(M)\,b(M)\,\tilde\delta_{\rm I}(M_{\rm c})\;
\label{barbm}
\end{equation}
and 
\begin{equation}\begin{split}
\bar{b}_{\delta_{\rm{I, sat}}} = \bar{n}_{\rm g}^{-1}\int_{M_{*}}^{\infty} &dM n(M)b(M)\,\int_{M_{*}}^{M} dm_{\rm sh} \frac{dN(m_{\rm sh}| M)}{dm_{\rm sh}} \\ &\times \tilde\delta_{\rm I}(m_{\rm sh}) u_{\rm d}(k,M)\;.
\end{split}
\end{equation}

\begin{figure}
\centering
\includegraphics[scale=0.45]{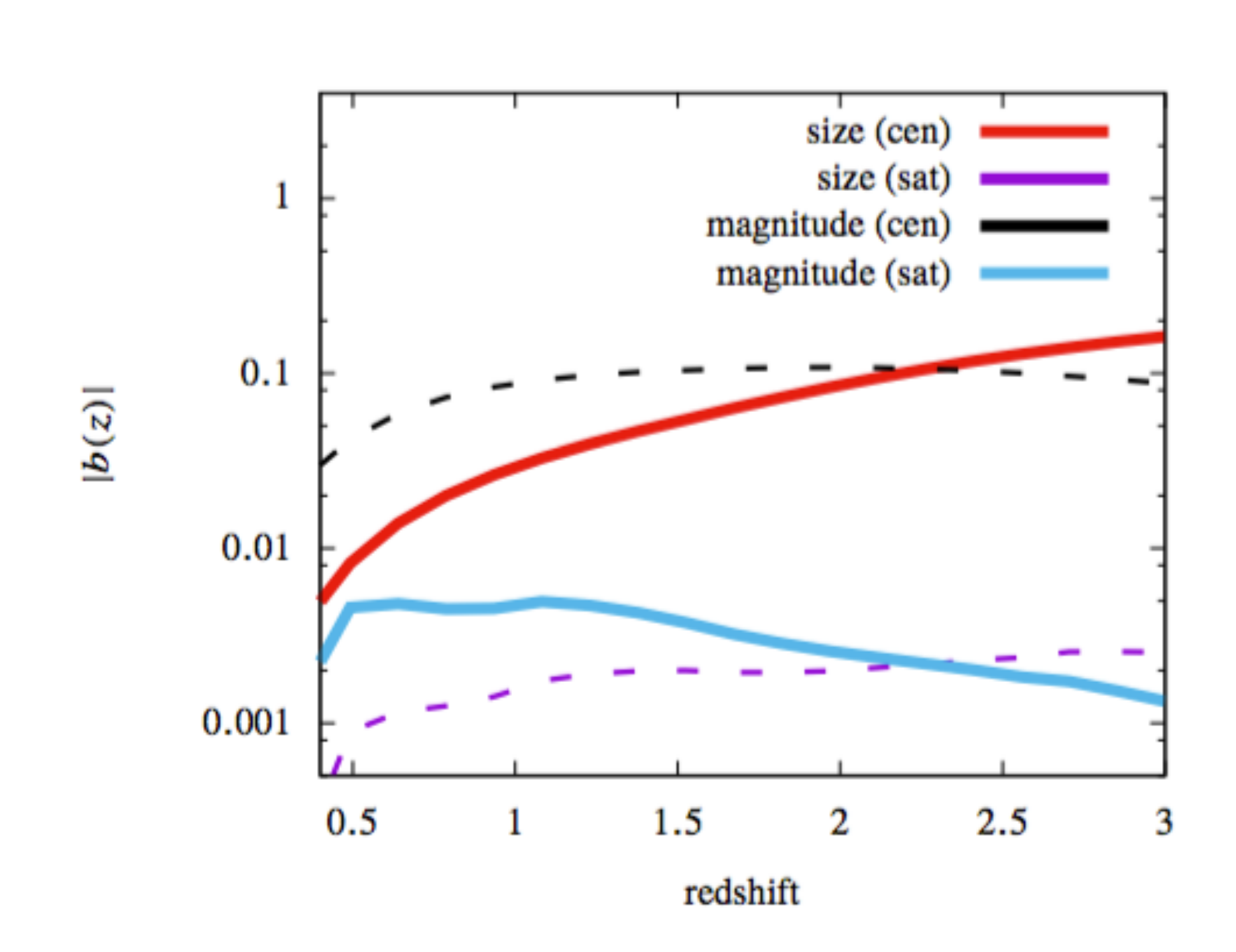}
\caption{Bias factors driving two-halo term for central and satellite galaxies in intrinsic size and magnitude correlations in Bj-band. Negative values are indicated with a dashed line while positive with a solid line.  At low redshifts, the magnitude biases tend to be higher than those of the sizes.}
\label{fig:BIAS-tot_Bj}
\end{figure}

As discussed in \citetalias{Ciarlariello2015}, it is these bias factors, particularly those for the central galaxies, that are most important in evaluating the importance of the intrinsic clustering for the interpretation of the weak lensing signal.  In Figure \ref{fig:BIAS-tot_Bj}, we show how these biases evolve in redshift.  As expected, the magnitude biases tend to have the opposite sign as the size biases, because larger densities will be correlated with higher luminosity and therefore smaller magnitudes.  The magnitude biases are also generally higher than the size biases, due to the steepness of the Vale-Ostriker mass-luminosity relation.  As a result, we expect the intrinsic correlations to be more of a contaminant for convergence estimation than was the case for sizes \citep{Ciarlariello2015}.

\subsection{Density-size cross power spectra}
For the cross-correlation density-size we obtain for both central and satellites:
\begin{equation}\begin{split}
P_{\rho\delta_{I}}^{\rm{1h-sat}} (k) &= \bar{\rho}^{-1}\bar{n}_{\rm g}^{-1} \int_{0}^{\infty} dM n(M) M  \\
&\times\int_{M_{*}}^{M} dm_{\rm sh}\frac{dN(m_{\rm sh}| M)}{dm_{\rm sh}} \tilde\delta_{\rm I}(m_{\rm sh})\,u(k, M)\, u_{\rm d}(k, M) \\
P_{\rho\delta_{\rm I}}^{\rm{2h} }(k) &= \bar{b}_{\rho}(\bar{b}_{\delta_{\rm{I, cen}}} +\bar{b}_{\delta_{\rm{I, sat}}}) P^{\rm{lin}}(k) 
\end{split}
\end{equation}
where $\bar{b}_{\rho}$ is the mass weighted bias defined in \citetalias{Ciarlariello2015} (using the prescriptions given in section (\ref{halomodel-sec})) and the other bias factors are given above. 

As in \citetalias{Ciarlariello2015}, we are assuming all of the lensing mass is associated with the haloes, and ignore mass associated with sub-clumps.  Potentially this approximation fails to take into account further correlations between size and density on scales within haloes but it should be good to understand the large-scale behaviour.

\subsection{Size-magnitude cross-power spectra}

Below, we will explore combined size and magnitude estimators for the convergence, which potentially can increase the signal-to-noise.  In addition, many of their systematics may be uncorrelated, making their cross-correlation measurements more robust.  However, the intrinsic size and magnitudes both correlate with the underlying density field, meaning the intrinsic correlations remain an important potential systematic.  We discuss the intrinsic size-magnitude cross spectra here. 

The one-halo terms for the size-magnitude cross-power spectrum are:
\begin{equation}\begin{split}
P^{\rm{1h-sat}}_{\delta_{\rm I}^{\lambda}\delta_{\rm I}^{\rm m}}(k) &= \bar{n}_{\rm g}^{-2} \int_{M_{*}}^{\infty} dM n(M)\\
&\times\left[\int_{M_{*}}^{M} dm \frac{dN(m_{\rm sh}| M)}{dm_{\rm sh}}\tilde\delta_{\rm I}^{\lambda}(m_{\rm sh}) u_{\rm d}(k, M)\right]\\
&\times\left[\int_{M_{*}}^{M} dm \frac{dN(m_{\rm sh}| M)}{dm_{\rm sh}}\tilde\delta_{\rm I}^{\rm m}(m_{\rm sh}) u_{\rm d}(k, M)\right]. \end{split}
\end{equation}

The contribution from central-satellite correlation terms is made of two parts, depending on whether we consider the size or the magnitude of the central galaxy:
\begin{equation}\begin{split}
P^{\rm{1h-cen-sat}}_{\delta_{\rm I}^{\lambda}\delta_{\rm I}^{\rm m}}&(k) = \frac{1}{\bar{n}_{\rm g}^2} \int_{M_{*}}^{\infty} dM n(M)\,  \\
&\times [\tilde\delta_{\rm I}^{\lambda}(M_{\rm c})\int_{M_{*}}^{M} dm_{\rm sh} \frac{dN(m_{\rm sh}| M)}{dm} \tilde\delta_{\rm I}^{\rm m}(m_{\rm sh}) u_{\rm d}(k, M) \\ &+ \tilde\delta_{\rm I}^{\rm m}(M_{\rm c})\int_{M_{*}}^{M} dm_{\rm sh} \frac{dN(m_{\rm sh}| M)}{dm} \tilde\delta_{\rm I}^{\lambda}(m_{\rm sh}) u_{\rm d}(k, M)].
\end{split}
\end{equation}

For the two-halo terms we have three contributions: from centrals, from satellites and from the central-satellite term:
\begin{equation}\begin{split}
&P_{\delta_{\rm{I}}^{\lambda}\delta_{\rm{I}}^{\rm m}}^{\rm{2h-cen}}(k) = \bar{b}_{\delta_{\rm{I,cen}}^{\lambda}}\bar{b}_{\delta_{\rm{I, cen}}^{\rm m}}P^{\rm lin}(k) \\
&P_{\delta_{\rm{I}}^{\lambda}\delta_{\rm{I}}^{\rm m}}^{\rm{2h-sat}}(k) = \bar{b}_{\delta_{\rm{I, sat}}^{\lambda}}\bar{b}_{\delta_{\rm{I, sat}}^{\rm m}}P^{\rm lin}(k) \\
&P_{\delta_{\rm{I}}^{\lambda}\delta_{\rm{I}}^{\rm m}}^{\rm{2h-cen-sat}}(k) = (\bar{b}_{\delta_{\rm{I, cen}}^{\lambda}}\bar{b}_{\delta_{\rm{I, sat}}^{\rm m}} + \bar{b}_{\delta_{\rm{I, cen}}^{\rm m}}\bar{b}_{\delta_{\rm{I, sat}}^{\lambda}})P^{\rm lin}(k).
\end{split}
\end{equation}

\section{Results}

In this section we discuss how the three-dimensional power spectra discussed above affect the projected two-dimensional lensing estimation for realistic surveys.  We focus on two cases: a single projected sample and a tomographic analysis of many redshift slices.  In the former case, the different redshift dependences of the size and magnitude responsivity functions mean that we cannot naively combine the two estimators.  

\subsection{Model assumptions}

We evaluate our results in the context of a flat $\Lambda$CDM cosmology with parameters consistent with first-year Planck data \citep{PlanckColl2014}; in particular, we assume a total matter density  $\Omega_{\rm{m,0}} = 0.32$, cosmological constant density $\Omega_{\Lambda,0} = 0.68$, baryon density  $\Omega_{\rm{b,0}} = 0.049$ and Hubble constant $H_{0} = 100\,{h}\,{\rm km}\,{\rm s}^{-1}\,{\rm Mpc}^{-1}$, where $h = 0.67$.  In addition, we assume the spectral index of the matter power spectrum is $n_{\rm s} = 0.96$ and it is normalised such that $\sigma_{8} = 0.83$.

For estimating lensing convergence power spectrum, we adopt the transfer function given in \cite{Eisenstein1998} and non-linear evolution of the matter power spectrum  is calculated with HALOFIT from \cite{Smith2003} recently revised by \cite{Takahashi2012}.  This non-linear treatment provides very comparable answers to our halo model calculations. 
 
In Section \ref{sizemagpdf} we noted that our halo modelling of the size-magnitude distribution leads to a realistic redshift distribution for a Euclid-like survey.  
However for simplicity we adopt the following commonly used parameterisation in order to calculate the angular power spectra: 
\begin{equation}
p(z) \propto z^{a}\exp{\left[-\left(\frac{z}{z_{0}}\right)^{b}\right]}\;,
\label{redsource}
\end{equation}
which is normalized to unity in order to have the number density distribution given by : $n(z) = \bar{n}p(z)$.
Following \cite{Schneider2010}, we consider a set of parameters to simulate a Euclid-like survey spanning 15000 square degrees with an average galaxy number density per steradian $\bar{n} = 30\,\,\rm{arcmin}^{-2}$ and we assume the parameters to be $a=2$, $b=1.5$, $z_{0}=0.64$ which gives a mean redshift around $0.96$.

\subsection{2D lensing}

First, we compare the importance of the intrinsic size and magnitude correlations to the lensing effects for a thick redshift slice.  We do this by considering directly the statistics of the projected size and magnitude fields, $\delta_{\rm I}^\lambda$ and $\delta_{\rm I}^m$, weighted by the redshift distribution $p(z)$:
\begin{equation}
\delta_{\rm I}(\theta) = \int d\chi\,p(\chi)\,\delta_{\rm I}(\chi\theta, \chi)\;.
\end{equation}
To compare with the relevant weak lensing power spectra, we must weight the convergence with the potentially redshift dependent responsivity factors $\eta_\lambda(z)$ and $\eta_m(z)$.    

\begin{figure}
\centering
\includegraphics[scale=0.45]{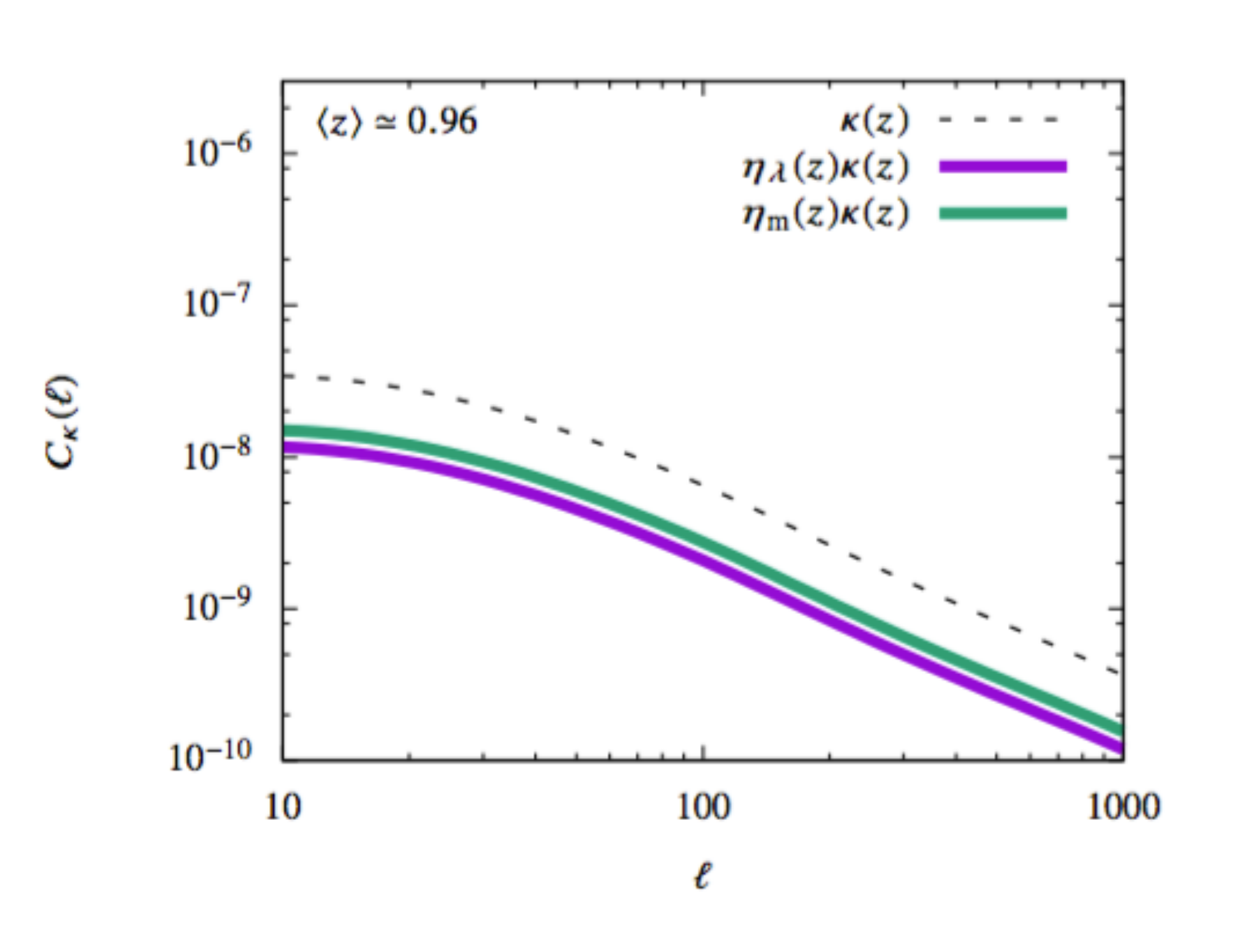}
\caption{Convergence power spectra for Euclid-like survey when redshift dependence of the responsivity values is included.}
\label{fig:EUCLID_convcl}
\end{figure}

We incorporate the responsivity factors  into an effective re-weighting of  the lensing convergence power spectrum $C_\kappa$, modifying the usual  Limber approximation \citep{Limber1954}:
 \begin{equation}
 C_{\kappa-{\rm eff}}(\ell) = \left(\frac{3 H_0^2 \Omega_{\rm{m,0}}}{2c^2}\right)^2\int_0^{\chi_{\rm{hor}}} d\chi \, 
\frac{g^{2}(\chi)}{[f_{K}(\chi)]^2} \, P_\delta \left( \frac{\ell}{f_{K}(\chi)},\chi \right)\;,
\label{convergence}
\end{equation}
where $P_{\delta}$ is the matter power spectrum, $\chi$ is the comoving distance along the line of sight, $\chi_{\rm{hor}}$ is the comoving horizon distance and  $c$ is the speed of light.  $f_{K}(\chi)$ is the comoving angular diameter distance that  depends on $K$, the inverse square of curvature radius in units of $H_{0}/c$,  as follows:
\begin{equation}
 f_{K}(\chi)=\left\{
      \begin{array}{ll}
        \sqrt{K}\sin(\sqrt{K}\chi) & \quad K>0 \\
      \chi & \quad K=0 \;, \\
      \sqrt{-K}\sinh(\sqrt{-K}\chi) & \quad K<0 .
      \end{array}
    \right.
\end{equation}
For simplicity, below we will assume $K=0$.

The modified weighting function is 
\begin{equation}
g(\chi) =\frac{f_{K}(\chi)}{a(\chi)}\int_\chi^{\chi_{\rm{hor}}}\, d\chi'\ \eta_{x}(\chi')p(\chi') 
\frac{f_{K}(\chi'-\chi)}{f_{K}(\chi')}\;,
\label{lensingweight}
\end{equation}
where $a$ is the dimensionless scale factor. 
Notice that here we included the responsivity function $\eta_{x}(\chi')$ in eq. (\ref{lensingweight}). This is because of the redshift dependence of the responsivity due to the cuts in the survey. The effect is a different weighting of the convergence power spectrum depending whether we are measuring size or magnitude correlations. This difference is shown in Fig. \ref{fig:EUCLID_convcl}.  Note that when the responsivities are redshift independent, the effective convergence spectrum is simply $C_{\kappa-{\rm eff}}(\ell)  =  \eta_\lambda^2 C_{\kappa}(\ell)$ or $C_{\kappa-{\rm eff}}(\ell)  =  \eta_m^2 C_{\kappa}(\ell)$.  However, in general the responsivity redshift dependence makes their relation less straight forward.

Our primary observables will be the two-point moments of the observed size or magnitude fluctuation fields; these have three contributions, the lensing (``GG") contribution, the intrinsic (`II") contribution, and their cross correlation.   
In spherical harmonic space, these can be written as 
\begin{equation}
C_{\hat{\delta}}(\ell) = C_{\kappa-{\rm eff}}(\ell) + C_{\rm II}(\ell) + C_{\rm GI}(\ell) + \frac{\sigma_{\rm intr}^2}{\bar{n}}
\label{cell}
\end{equation}
where we have included a shot-noise term, $\sigma_{\rm intr}^2$ is the variance of the total intrinsic sizes or magnitudes and $\bar{n}$ is the total number of galaxies per steradian.
 For both sizes and magnitudes, we have that the II and the GI terms are:
\begin{equation}\begin{split}
C_{\rm II}(\ell) &= C_{\delta_{\rm I}}(\ell) \\
C_{\rm GI}(\ell) &= 2C_{\kappa-{\rm eff}\delta_{\rm I}}(\ell).
\end{split}
\end{equation}

Again we use Limber's approximation to calculate the intrinsic terms (for both size and magnitude) for the II and GI terms as follows:
\begin{equation}\begin{split}
&C_{\delta_{\rm I}^{\lambda}}(\ell) =  \int_0^{\chi_{\rm{hor}}} d\chi \, 
\frac{p^{2}(\chi)}{\chi^2} \, P_{\delta_{\rm I}^{\lambda}}\left( \frac{\ell}{\chi},\chi \right) \\
&C_{\delta_{\rm I}^{\rm m}}(\ell) =  \int_0^{\chi_{\rm{hor}}} d\chi \, 
\frac{p^{2}(\chi)}{\chi^2} \, P_{\delta_{\rm I}^{\rm m}}\left( \frac{\ell}{\chi},\chi \right) \\
&C_{\kappa-{\rm eff}\delta_{\rm I}^{\lambda}}(\ell) = \frac{3 H_0^2 \Omega_{\rm{m,0}}}{2c^2}\int_0^{\chi_{\rm{hor}}} d\chi \, 
\frac{g(\chi)p(\chi)}{\chi^2} \, P_{\rho\delta_{\rm I}^{\lambda}} \left( \frac{\ell}{\chi},\chi \right) \\
&C_{\kappa-{\rm eff}\delta_{\rm I}^{m}}(\ell) = \frac{3 H_0^2 \Omega_{\rm{m,0}}}{2c^2}\int_0^{\chi_{\rm{hor}}} d\chi \, 
\frac{g(\chi)p(\chi)}{\chi^2} \, P_{\rho\delta_{\rm I}^{\rm m}} \left( \frac{\ell}{\chi},\chi \right).
\end{split}
\label{angularlimber}
\end{equation}

\begin{figure}
\centering
\includegraphics[scale=0.3]{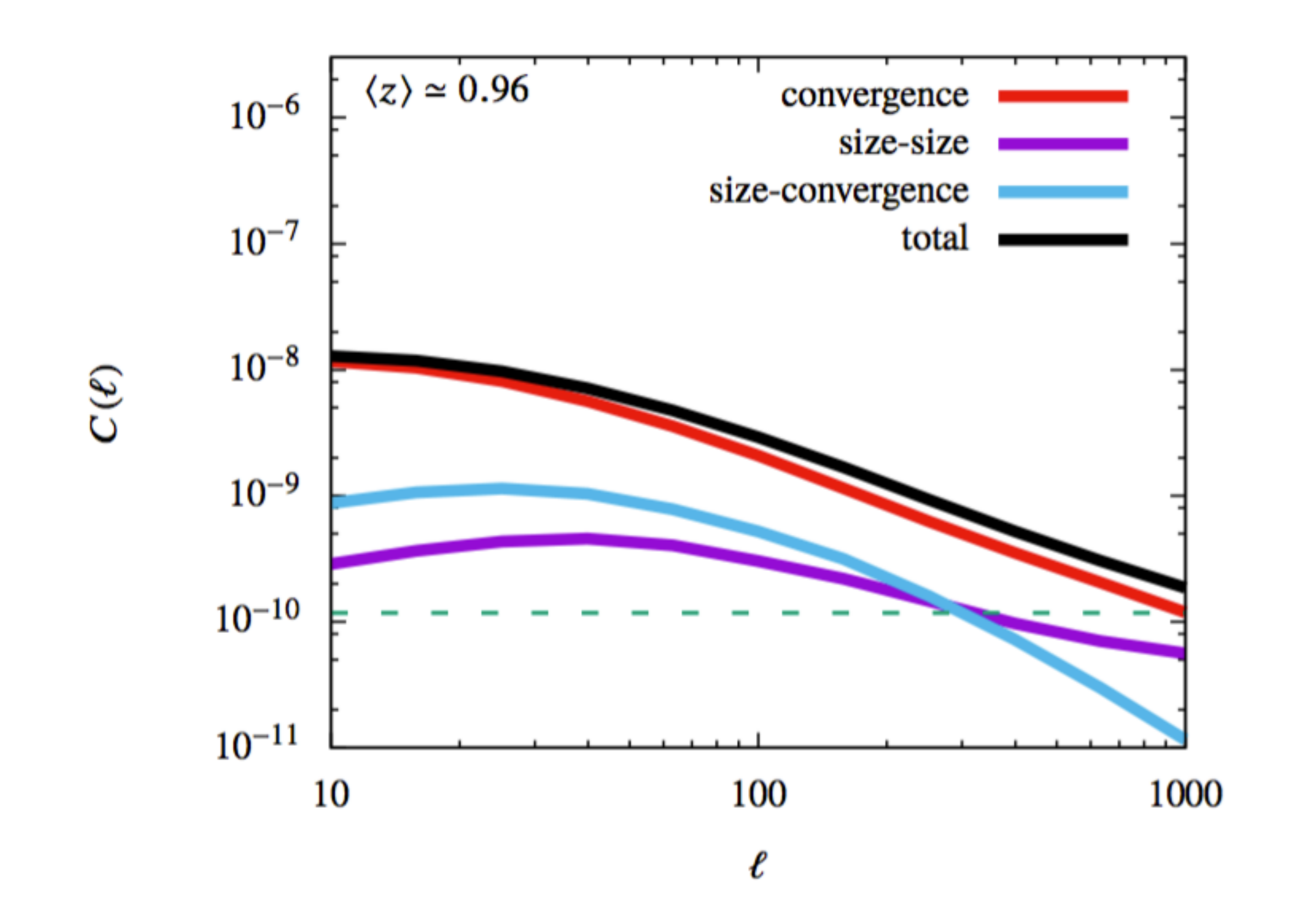}
\caption{2D-only size power spectra in $\rm{B}_{\rm j}$-band for a Euclid-like survey. Shot-noise is shown separately (dashed line) .}
\label{fig:RAD_Bj}
\end{figure}

\begin{figure}
\centering
\includegraphics[scale=0.3]{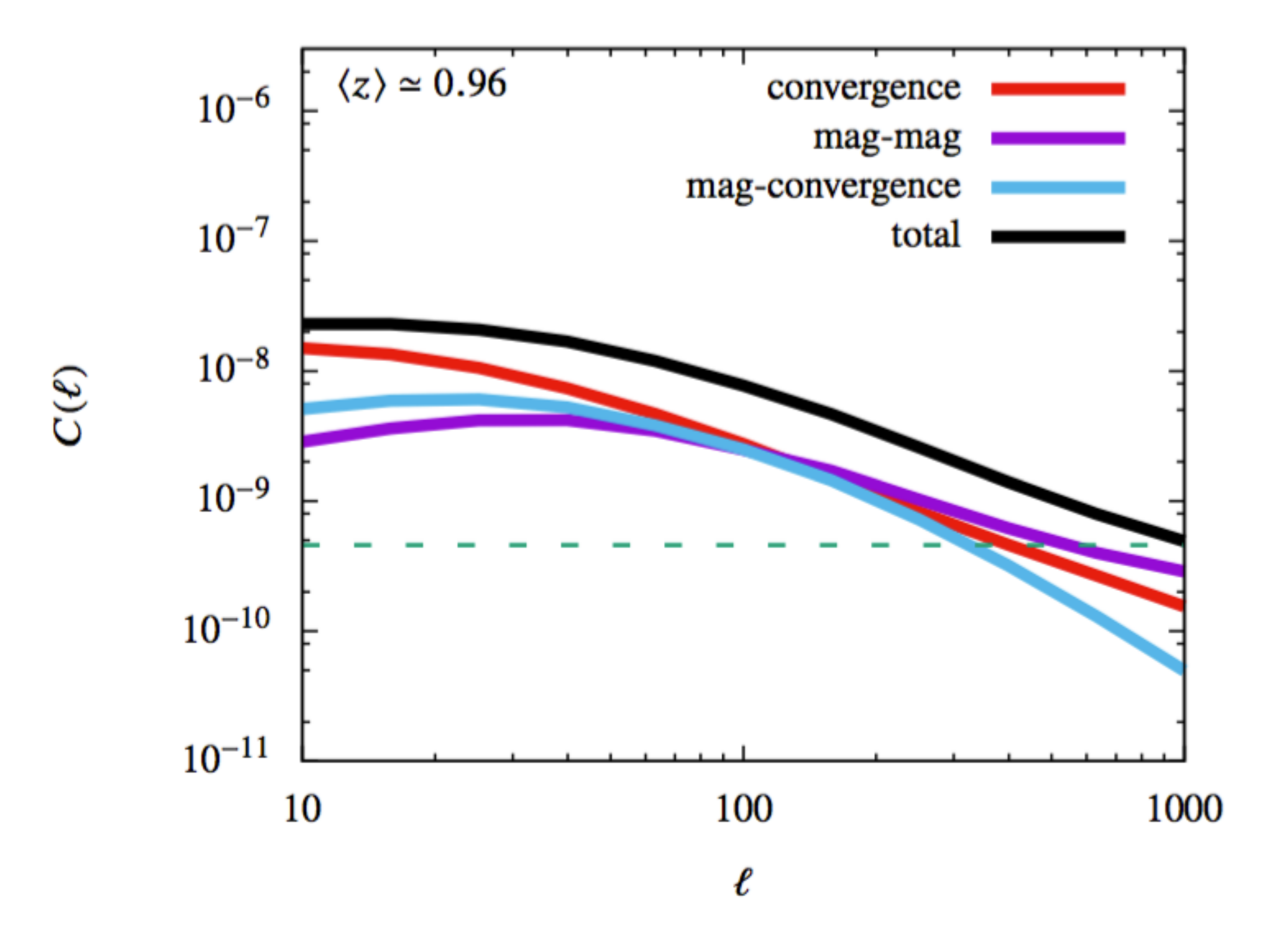}
\caption{2D-only magnitude power spectra in $\rm{B}_{\rm j}$-band for a Euclid-like survey. Shot-noise is also shown separately (dashed line).}
\label{fig:MAG_Bj}
\end{figure}

The full projected power spectra can be seen in Figures \ref{fig:RAD_Bj} and \ref{fig:MAG_Bj} for the sizes and magnitudes, respectively.  In each figure, we plot the three (GG, II and GI) contributions to the total observed spectrum.  
Previously \citetalias{Ciarlariello2015} analysed the ideal case, with a fixed mass cutoff at all redshifts and ideal responsivities. 
Figure \ref{fig:RAD_Bj} largely confirms the findings of  \citetalias{Ciarlariello2015} for more realistic observations: for a reasonably deep redshift survey the purely intrinsic correlations appear to comprise a minor contribution to the overall signal, while the cross correlation between the intrinsic and lensing terms does introduce a 10\% contribution.  While subdominant, this is enough to potentially bias the pure lensing interpretation of the observed correlations.    

Figure \ref{fig:MAG_Bj} extends this analysis to the magnitude fluctuation spectrum.  
The level of contamination from intrinsic correlations is higher, consistent with the enhancement of  seen for the magnitude bias.  On large angular scales the convergence is still dominant, but on small angular scales the intrinsic II signal is comparable and it even dominates for higher multipoles. 

\subsubsection{Comparison with Intrinsic Alignments}

\begin{figure}
\centering
\includegraphics[scale=0.3]{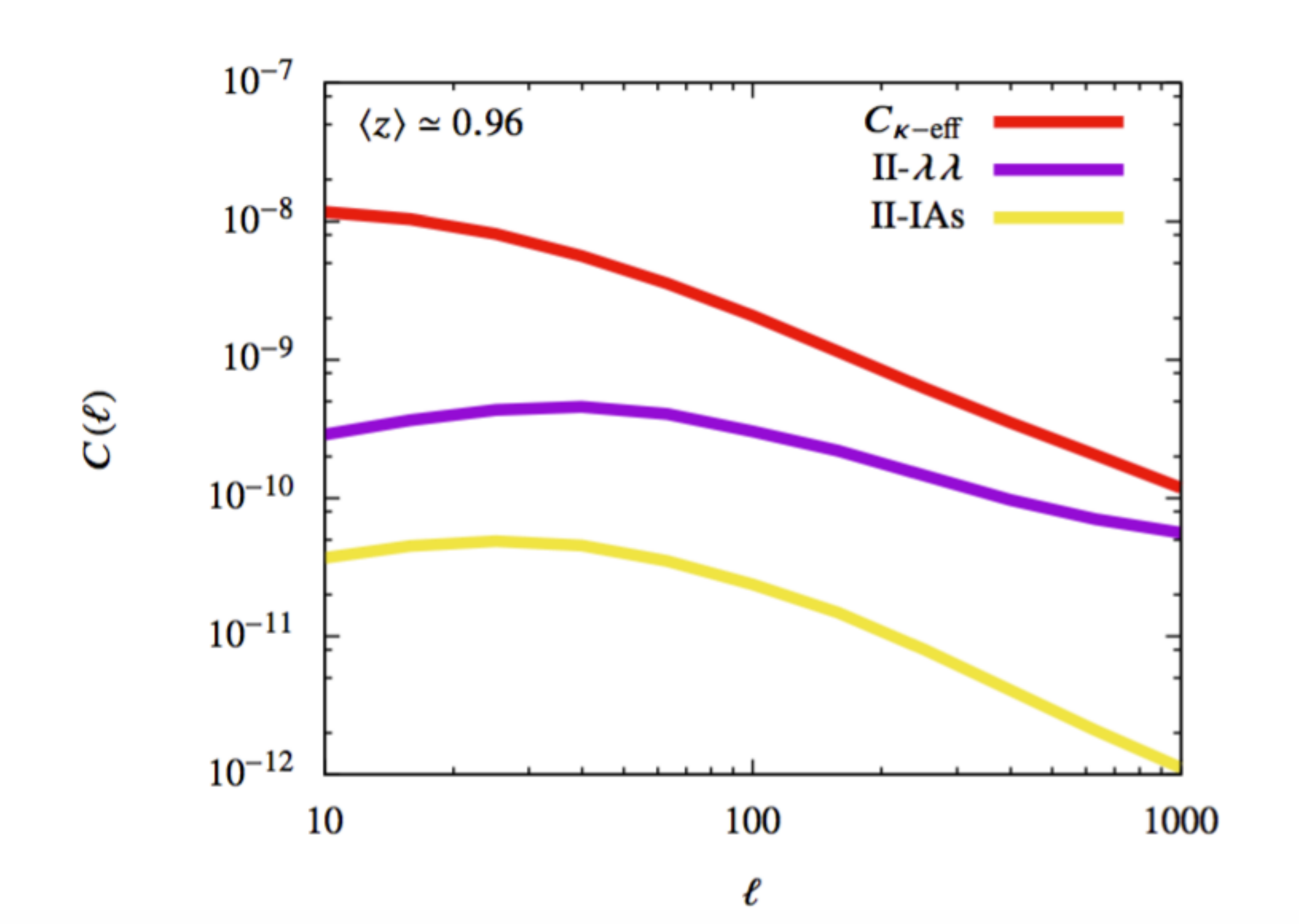}
\caption{Comparison of the contamination due to intrinsic size correlations (II-$\lambda\lambda$) and intrinsic alignments (II-IAs) with respect to the convergence signal ($C_{\kappa-\rm{eff}}$) expected in a Euclid-like survey.}
\label{fig:ias2d}
\end{figure}

\begin{figure}
\centering
\includegraphics[scale=0.3]{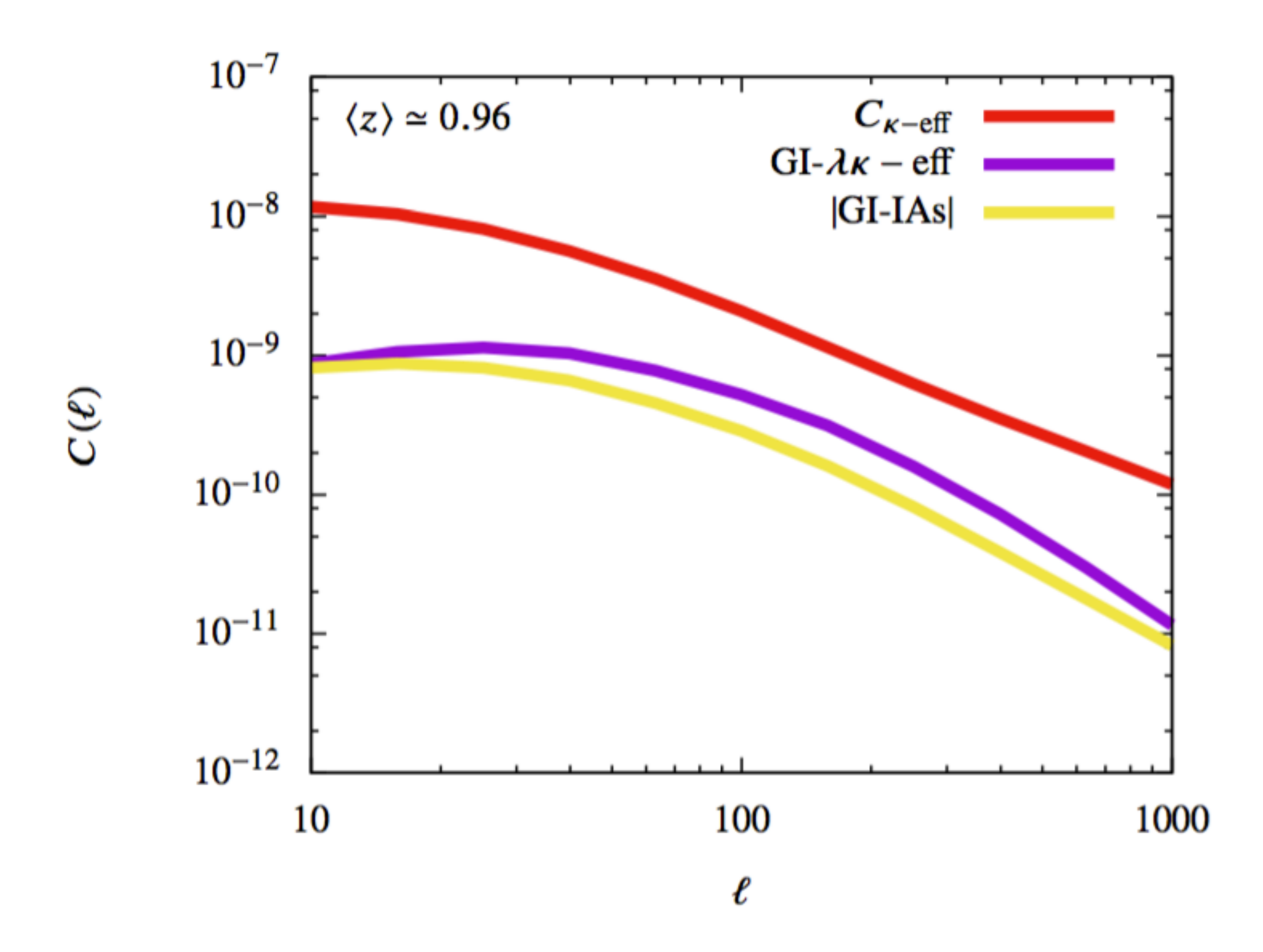}
\caption{Comparison of the contamination due to intrinsic size-convergence correlations (GI-$\lambda\lambda$) and intrinsic alignments (GI-IAs) with respect to the convergence signal ($C_{\kappa-\rm{eff}}$) expected in a Euclid-like survey.}
\label{fig:ias2d_gi}
\end{figure}

It is useful to understand how the intrinsic size and magnitude correlations in magnification measurements 
compare as a potential systematic to the importance of intrinsic alignments for shear measurements.   
In Figs. \ref{fig:ias2d}  and \ref{fig:ias2d_gi}  we calculate the expected size and ellipticity correlations, focussing on the II and GI contributions respectively, and compare them to the convergence power spectrum expected for a Euclid-like survey (as in Fig. \ref{fig:RAD_Bj}). The ellipticity term is calculated following the fiducial intrinsic alignment model described in eq. (29) of \cite{Alsing2015}, itself following the normalisation of \cite{Bridle2007} and \cite{Kirk2012}. 
For a comparison on small scales, we apply the so-called non-linear alignment (NLA) model (e.g.\cite{Troxel2015}), using the non-linear evolution of the matter power spectrum calculated with HALOFIT from \cite{Smith2003} and revised by \cite{Takahashi2012}.

We see that while both systematics are sub-dominant with respect to the convergence power spectrum, our estimate of the intrinsic size-size term is somewhat larger than the intrinsic alignment correlation.    This is particularly true for the II contribution (Fig. \ref{fig:ias2d}), while the GI contributions are more comparable (Fig. \ref{fig:ias2d_gi}).   The latter is more relevant when considering their impact as a systematic, as II contaminations can be reduced by correlating galaxies separated in redshift.    While the somewhat larger intrinsic size correlations might suggest larger biases in the interpretation of size measurements, this will depend sensitively on the quality of the modelling, the number of degrees of freedom to be marginalised over, and how well these can be constrained by independent measurements, such as the galaxy size-mass relation or direct cross-correlation of the large-scale size and density fields.  We leave a 
more in-depth study of the impact of intrinsic size correlations on cosmological parameter inference for future study. 

\subsection{Lensing tomography}
 
With weak lensing tomography, described for example in \cite{Hu1999}, we can increase what we learn about cosmology by subdividing the survey into a number of photometric redshift slices.  This has a number of advantages, particularly in understanding the evolution of the convergence field.  It also has the benefit of localising the II correlations, which dominate on short distances.  Further, in thinner bins, we are able to treat the responsivity factors as effectively constant, making combining size and magnitude estimators of the convergence simpler. 

 \subsubsection{Separate size and magnitude estimators}
We divide the redshift distribution given in eq. (\ref{redsource})  into $N_{\rm bin}=5$ tomographic bins of width $\Delta z_{\rm bin}=0.3$, cutting at $z_{\rm max} =1.5$, such that:
\begin{equation}
p^{i}(z) = \left\{
      \begin{array}{ll}
        p(z) & \quad z_{i}<\,z\,\leq z_{i+1} \\
      0 & \quad \mbox{otherwise} \\
      \end{array}
    \right.
\end{equation}
for every $i=1,\dots,N_{\rm bin}$ and $z_{i}$ and $z_{i+1}$ indicate the bin boundaries; we normalise such that:
\begin{equation}
\int_{0}^{z_{\rm max}} dz\,p^{i}(z) = 1.
\end{equation}
For simplicity we do not include any photometric redshift error. This is potentially an important issue when considering intrinsic correlations in weak lensing; however, given the relatively coarse binning we are considering here, we expect these corrections to be small.  Finer binning might allow for better control of the intrinsic correlations, but then the photometric errors would require a more careful treatment. 

Measuring the correlations in narrow redshift bins, where the responsivities can be considered constant, allows us to combine size and magnitude information because the responsivities can be taken out of the redshift integration. We can construct estimators directly for the convergence over a given slice using:
\begin{equation}\begin{split}
&\hat{\kappa}^{\lambda} \equiv  \frac{1}{{\eta}_{\rm \lambda}}(\lambda - \langle \lambda \rangle)\;, \\
&\hat{\kappa}^{\rm m} \equiv  \frac{1}{{\eta}_{\rm m}}(m - \langle m \rangle)\;.
\end{split}
\end{equation}
Using  $i$ and $j$ to represent two different redshift bins, the equations from previous sections are then slightly modified:
\begin{equation}\begin{split}
C^{ij}_\kappa(\ell) = & \left(\frac{3 H_0^2 \Omega_{\rm{m,0}}}{2c^2}\right)^2 \int_0^{\chi_{\rm{hor}}} d\chi \, 
\frac{g^i(\chi)g^{j}(\chi)}{[f_{K}(\chi)]^2} \, P^{ij}_\delta \left( \frac{\ell}{f_{K}(\chi)},\chi \right) \\
C^{ij}_{\kappa_{\rm I}^{\lambda}}(\ell) =&  \frac{1}{\eta_{\lambda}^{i}\eta_{\lambda}^{j}}\int_0^{\chi_{\rm{hor}}} d\chi \, 
\frac{p^{i}(\chi)p^{j}(\chi)}{\chi^2} \, P^{ij}_{\delta_{\rm I}^{\lambda}}\left( \frac{\ell}{\chi},\chi \right) \\
C^{ij}_{\kappa_{\rm I}^{m}}(\ell) =&  \frac{1}{\eta_{m}^{i}\eta_{m}^{j}}\int_0^{\chi_{\rm{hor}}} d\chi \, 
\frac{p^{i}(\chi)p^{j}(\chi)}{\chi^2} \, P^{ij}_{\delta_{\rm I}^{\rm m}}\left( \frac{\ell}{\chi},\chi \right) .
\end{split}
\end{equation}
Here, the tomographic lensing kernel for a given bin is defined as usual,
\begin{equation}
g^{i}(\chi) =  \frac{f_{K}(\chi)}{a(\chi)}\int_{\chi}^{\chi_{\rm hor}}\, d\chi'\ p^i(\chi')
\frac{f_{K}(\chi'-\chi)}{f_{K}(\chi')}.
\label{lensingweight-bin}
\end{equation}
Since the redshift bins are not overlapping and photometric redshift errors are not included, the II term vanishes in bin cross-correlation and contributes only when $i=j$; for the GI contributions, the term $g^{i}(\chi)p^{j}(\chi)$ is zero when considering $i<j$.

We also have the GI contributions to the total correlations, which are:
\begin{equation}\begin{split}
C^{ij}_{\kappa\kappa_{\rm I}^{\lambda}}(\ell) =& \frac{3 H_0^2 \Omega_{\rm{m,0}}}{2c^2}\frac{1}{\eta_{\lambda}}\int_0^{\chi_{\rm{hor}}} d\chi \, 
\frac{p^{i}(\chi)g^{j}(\chi) + g^{i}(\chi)p^{j}(\chi)}{\chi^2} \\
 & \times\,P^{ij}_{\kappa\delta_{\rm I}^{\lambda}} \left( \frac{\ell}{\chi},\chi \right) \\
C^{ij}_{\kappa\kappa_{\rm I}^{m}}(\ell) =& \frac{3 H_0^2 \Omega_{\rm{m,0}}}{2c^2}\frac{1}{\eta_{\rm m}}\int_0^{\chi_{\rm{hor}}} d\chi \, 
\frac{p^{i}(\chi)g^{j}(\chi) + g^{i}(\chi)p^{j}(\chi)}{\chi^2} \\
 & \times\,P^{ij}_{\kappa\delta_{\rm I}^{m}} \left( \frac{\ell}{\chi},\chi \right) .
\end{split}
\end{equation}
Finally to combine size and magnitude estimators, we must also calculate the cross-correlation between their intrinsic correlations: 
\begin{equation}
C^{ij}_{\kappa_{\rm I}^{\lambda}\kappa_{\rm I}^{\rm m}}(\ell) = \frac{1}{\eta_{\lambda}^i\eta_{m}^j}\int_0^{\chi_{\rm{hor}}} d\chi \, 
\frac{p^{i}(\chi)p^{j}(\chi)}{\chi^2} \, P^{ij}_{\delta_{\rm I}^{\lambda}\delta_{\rm I}^{\rm m}} \left( \frac{\ell}{\chi},\chi \right).
\label{angulartom}
\end{equation}

\begin{figure*}
\centering
\includegraphics[scale=0.6]{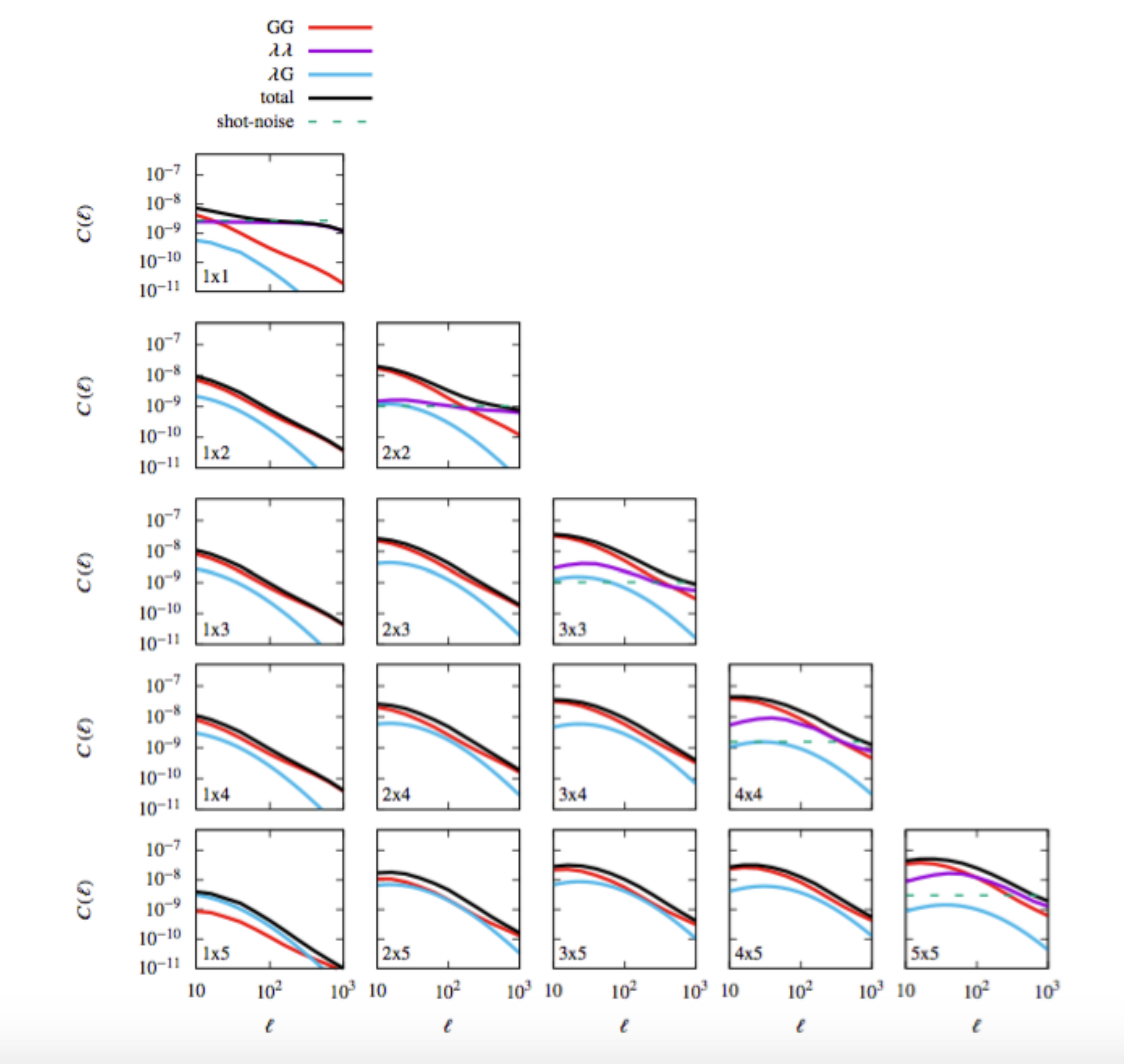}
\caption{The tomographic power spectra in $\rm{B}_{\rm j}$-band for a Euclid-like survey using the size estimator. 
Each redshift bin has $\Delta\,z=0.3$ (Bin 1 corresponds to the range from $z=0$ to $z=0.3$.) 
Responsivity values are different in each bin, as indicated in Table \ref{tb:values}.  
Lensing convergence is indicated with GG and the II term only contributes to diagonal correlations. Shot-noise is shown by the dashed lines.  }
\label{fig:RAD-tom-Bj}
\end{figure*}

\begin{figure*}
\centering
\includegraphics[scale=0.6]{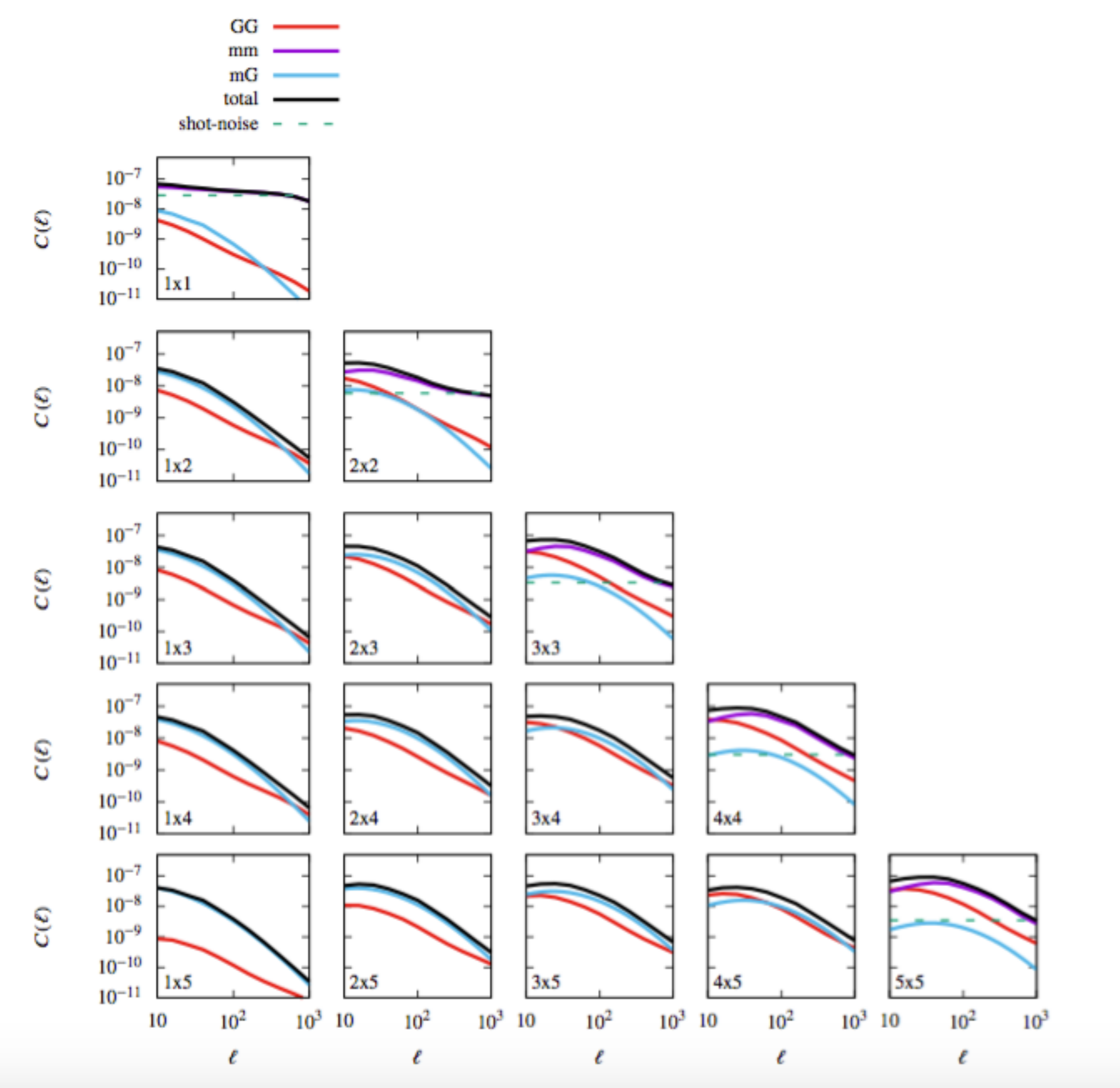}
\caption{The tomographic power spectra in $\rm{B}_{\rm j}$-band for a Euclid-like survey using the magnitude estimator. 
Responsivity values are different in each bin, as indicated in Table \ref{tb:values}.  
Lensing convergence is indicated with GG and the II term only contributes to diagonal correlations. Shot-noise is shown by the dashed lines. 
}
\label{fig:MAG-tom-Bj}
\end{figure*}

The results of the individual size and magnitude tomographic analyses are shown respectively in Figures \ref{fig:RAD-tom-Bj} and \ref{fig:MAG-tom-Bj}.
For the size estimator, we see that the convergence dominates the estimator in most correlations.  
The auto-correlations include the II terms, and these are comparable to the lensing signal at small angular scales and at low redshifts, where the lensing is signal is small.  However, in these situations the signals tend to be below the shot noise level.  In the cross correlations, the GI correlations are usually 
small, with the exception of the $1\times 5$ correlation, where the convergence is small. 

In the case of the magnitude estimator (Fig. \ref{fig:MAG-tom-Bj}), the intrinsic correlations are a much more serious systematic, again due to their larger bias.  The auto-correlations are  dominated by the intrinsic II signal even for the highest redshift bins.  In the cross-correlations, the GI signal is usually comparable to that of the convergence.

\begin{table}
\begin{center}
\begin{tabular}{| c | c | c | c | c | c | }
\hline
  $z_{\rm min}: z_{\rm max}$ & $\sigma_{\lambda}$  & $\sigma_{m}$ & $\sigma_{\lambda m}^2$ & $\alpha_{\lambda}$ & $\alpha_{m}$  \\
\hline
  $[0.0:0.3]$ & 0.391 & 1.092  & -0.103  & 0.977 & 0.023 \\
\hline
  $[0.3:0.6]$ & 0.438 & 1.040  & -0.164  & 0.973 & 0.027 \\
\hline
  $[0.6:0.9]$ & 0.453 & 0.900  & -0.161  & 0.898 & 0.102  \\
\hline
  $[0.9:1.2]$ & 0.456 & 0.738  & -0.126  & 0.745 & 0.255  \\
\hline
  $[1.2:1.5]$ & 0.455 & 0.603  & -0.092  & 0.561 & 0.439 \\
\hline
\end{tabular}
\caption{Table with rms values for size and magnitudes in different redshift bins, and their implications for the weights of the noise-weighted combined estimator.}
 \label{tb:values}
 \end{center}
\end{table}

\subsubsection{Combining size and magnitude estimators}

\begin{figure*}
\centering
\includegraphics[scale=0.6]{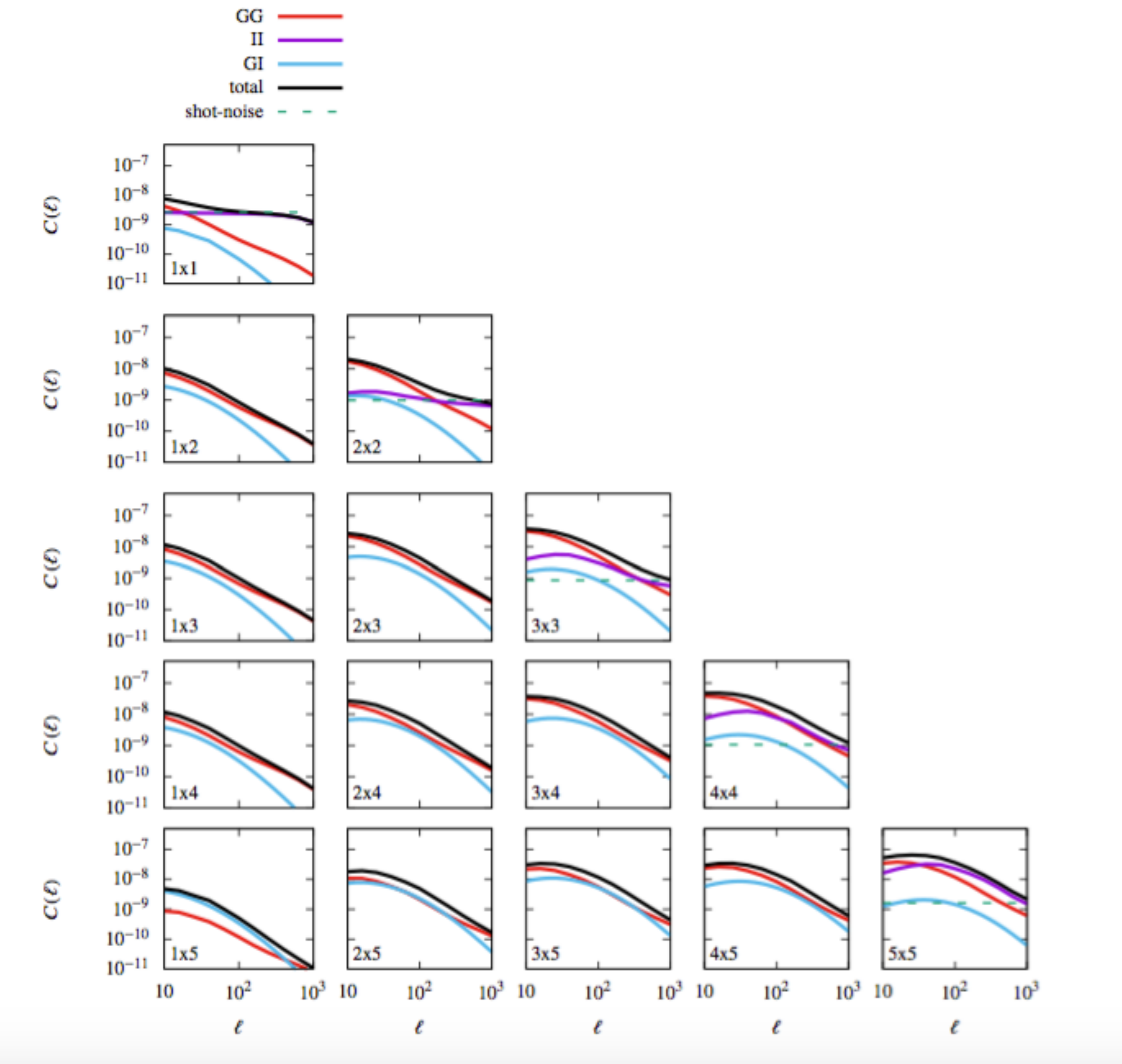}
\caption{
The tomographic power spectra in $\rm{B}_{\rm j}$-band for a Euclid-like survey using the noise-weighted combined estimator. 
The size and magnitude weightings are different in each bin, as indicated in Table \ref{tb:values}.  
Lensing convergence is indicated with GG and the II term only contributes to diagonal correlations. Shot-noise is shown by the dashed lines. 
}
\label{fig:EUCLID-tom-Bj}
\end{figure*}

A simple unbiased  noise-weighted estimator that combines the two estimators for galaxy sizes and magnitudes has already been presented in \cite{Schmidt2012} for the surface density and \cite{Alsing2015} for the convergence. Following their notation, we can write the total intrinsic contribution to the convergence as linear combination of two terms:
\begin{equation}
\hat{\kappa}_{\rm comb} = \alpha_{\lambda}\hat{\kappa}^{\lambda} + \alpha_{\rm m} \hat{\kappa}^{\rm m},
\end{equation}
where the two coefficient have the constraint $\alpha_{\lambda} + \alpha_{\rm m} = 1$ in order to have an unbiased estimator for the convergence.
Minimising the variance of the full estimator given above, we obtain explicit forms for $\alpha_{\lambda}$ and $\alpha_{m}$:
\begin{equation}\begin{split}
\alpha_{\lambda} &= \frac{\eta_{\lambda}^2\sigma_{\rm m}^2 - \eta_{\lambda}\eta_{\rm m}\sigma_{\lambda m}^2}{\eta_{\rm m}^2\sigma_\lambda^2 + \eta_{\lambda}^2\sigma_m^2 - 2\eta_{\lambda}\eta_{\rm m}\sigma_{\lambda m}^2} \\
\alpha_{m} &= \frac{\eta_{\rm m}^2\sigma_\lambda^2 - \eta_{\lambda}\eta_{\rm m}\sigma_{\lambda m}^2}{\eta_{\rm m}^2\sigma_\lambda^2 + \eta_{\lambda}^2\sigma_m^2 - 2\eta_{\lambda}\eta_{\rm m}\sigma_{\lambda m}^2}.
\end{split}
\end{equation}
Here the responsivities and the variances are calculated given a model (or observations) of the normalised size-magnitude distribution in a given bin:
\begin{equation}
\begin{split}
\sigma&_{\lambda}^2 = \int_{\lambda_{\rm lim}}^{+\infty} \int_{-\infty}^{m_{\rm lim}} d\lambda\,dm\,f(\lambda, m)(\lambda - \langle \lambda\rangle)^2 \\
\sigma&_{m}^2 = \int_{\lambda_{\rm lim}}^{+\infty} \int_{-\infty}^{m_{\rm lim}} d\lambda\,dm\,f(\lambda, m)(m - \langle m\rangle)^2 \\
\sigma&_{\lambda m}^2 = \int_{\lambda_{\rm lim}}^{+\infty} \int_{-\infty}^{m_{\rm lim}} d\lambda\,dm\,f(\lambda, m)(\lambda - \langle \lambda\rangle)(m - \langle m\rangle).
\end{split}
\end{equation}
For our model of the size-magnitude distribution, we find the values given in Table \ref{tb:values}.  The magnitude variance is larger at small redshifts, while the size variance changes little,  reflecting the trends that can be seen in Fig. \ref{fig:MATRIX_Bj}.
This gives more weight the size estimator at low redshift, and more equal weighting at higher redshifts. 

This full convergence estimator is unbiased, so the expected convergence power spectrum is unchanged;  
the II term is again diagonal in $i$ and $j$:
\begin{equation}\begin{split}
C^{ii}_{\rm II}(\ell) = & \alpha_{\lambda}^{i2} \left(C^{ii}_{\kappa_{\rm I}^{\lambda}}(\ell) + \frac{\sigma_{\lambda,\rm{intr}}^2}{\bar{n}^i\eta^{i2}_\lambda}\right)+ 
 \alpha_{\rm m}^{i2}\left(C_{\kappa_{\rm I}^{\rm m}}(\ell) + \frac{\sigma_{m,\rm{intr}}^2}{\bar{n}^i\eta^{i2}_m}\right)
 \\ & + 2\alpha_{\lambda}^{i}\alpha_{\rm m}^{i}C^{ii}_{\kappa_{\rm I}^{\lambda}\kappa_{\rm I}^{\rm m}}(\ell) , 
\end{split}
\end{equation}
where we have explicitly incorporated the separate shot noise contributions.  
For the GI term, we have:
\begin{equation}
C_{\rm GI}^{ij}(\ell) =\alpha_{\lambda}^{i}C^{ij}_{\kappa\kappa_{\rm I}^{\lambda}} + \alpha_{\lambda}^{j}C^{ji}_{\kappa\kappa_{\rm I}^{\lambda}}(\ell) + \alpha_{\rm m}^{i}C^{ij}_{\kappa\kappa_{\rm I}^{\rm m}}(\ell) + \alpha_{\rm m}^{j} C^{ji}_{\kappa\kappa_{\rm I}^{\rm m}}(\ell)
\end{equation}
and again, these terms will be zero for $i>j$.  

The results for the combined tomographic estimation are shown in Figure \ref{fig:EUCLID-tom-Bj}.  The shot noise levels are somewhat reduced with respect to their individual levels for size or magnitude.  
At low redshifts, the results are very comparable to the size estimator alone (Fig. \ref{fig:RAD-tom-Bj}) because the magnitude contributions are suppressed by their greater variance.     
Again, while the noise levels have dropped, the convergence dominates the auto-correlations where the signal is detectable. 
The cross-correlations are also largely dominated by the convergence signal, though the GI terms are somewhat higher than in the size-only estimator.  
The noise-weighted combination thus reduces the shot noise levels, while increasing the intrinsic correlation contamination.  

\section{Conclusions}

In our previous paper  \citep{Ciarlariello2015}, we examined the issue of intrinsic size correlations in a halo model, where the sizes of galaxies were assumed to be a simple function of the sub-halo mass.  Here, we have extended this analysis by examining the correlations in galaxy brightness, and by introducing intrinsic scatter in the mass-size and mass brightness relations.  We have also included realistic selection effects into our predictions to account for the reduced responsivity of the mean properties of galaxies to convergence. 

Overall, we find these improvements in the modelling have not affected the main conclusion of  \citet{Ciarlariello2015}, that intrinsic correlations in the galaxy properties used to trace magnification are an important systematic to measurements of the convergence power spectrum; if ignored, they can significantly bias the cosmological interpretation of the convergence measurements.   

The principle determining factor of the importance of the intrinsic correlations is their estimator weighted bias\citep{Ciarlariello2015}, e.g. Eq. \ref{barbm}.  These depend significantly on the form of the mass-size and mass-luminosity relations.
Because of the steeper relationship between the sub-halo mass and the luminosity reflected in the Vale and Ostriker (2008) relation, we expect a higher bias for the magnitude correlations compared to that expected for sizes, as can be seen in Figure \ref{fig:BIAS-tot_Bj}.   Because of this, the intrinsic contamination to magnitude correlations can actually be comparable to the convergence signal itself (Fig. \ref{fig:MAG_Bj}).   

Our results in have also been evaluated using a specific halo model, in particular using the \citet{Sheth1999} mass function and its associated halo bias model. If we instead use the \citet{Press1974} mass function and halo bias model can result in an increase of around 20\% in both size and magnitude bias. Thus, while there is some sensitivity to the implementation of the halo model, our main conclusions in terms of the impact of intrinsic size and magnitude correlations are not significantly affected. 

The addition of scatter in the mass-size and mass-luminosity relations does not directly affect expectations of the two-point correlations.  However, it does impact the probability that a galaxy of a given mass will be selected, and therefore the bias weighting of the sample.  Our modelling of the distribution indicates a significant sensitivity to the size and magnitude cuts, consistent with observations and indicating a responsiivity of the means to magnification significantly suppressed compared to the ideal values.  

In the absence of intrinsic systematics, it is beneficial to combine sizes and magnitudes together into a single noise-weighted estimator \citep{Alsing2015}.  However, given the difference in the expected intrinsic correlations, combining sizes and magnitudes may make the systematic contamination worse than for sizes alone.  But this may be mitigated depending on how the intrinsic correlations are marginalised over. 

The tomographic analysis shows that, like intrinsic shape correlations, intrinsic size and brightness correlations are a serious problem within narrow bins, and ameliorating them requires exploiting cross-correlations between bins where the II contributions are negligible.  However, at low redshifts, and in neighbouring bins, the GI terms can also be a comparable systematic.  

Our theoretical results emphasise the need to better quantify these intrinsic correlations, particularly on small scales where the halo model is approximate and potentially is missing important physics.  Hydrodynamic simulations have more realistic small scale physics, but may not have the full dynamic range essential for weak lensing analyses.   Semi-analytic models, based on simulated merging trees and constrained to match related galaxy observables, may improve the situation.  

Equally essential is to focus on measuring these effects in large scale surveys, focusing on low redshifts and large scales where the intrinsic signal is expected to dominate over shot noise and the convergence signal.  We are presently investigating whether these correlations can be observed in the SDSS.  Measurements of such correlations are observationally challenging and they are subject to many of the same systematics as shape measurements.   However, unlike shape estimators, the magnification estimators have the additional complication of requiring accurate measurements of the mean sizes and magnitudes and their responsivities to lensing under the selection function.  

\section*{Acknowledgements}

This work was advanced in part through a 2016 workshop on Cosmology with Size and Flux Measurements held at the International Space Science Institute in Bern. 

We thank David Bacon and the participants of the ISSI workshop, particularly Alan Heavens and Fabian Schmidt, for useful conversations.    RC is supported by the UK Science and Technologies Facilities Council grant ST/K00090X/1.  




\bibliographystyle{mnras}
\bibliography{CiarlarielloCrittenden_v2.bbl} 








\bsp	
\label{lastpage}
\end{document}